\documentclass[preprint]{aastex}

\def\arcs{\char'175\ }
\def\arcsc{\char'175 }
\def\arcm{\char'023\ }
\def\arcmn{\char'023 }

\def\etal{et~al.\ }

\def\hub{\ifmmode H_\circ\else H$_\circ$\fi}
\def\kms{~km~s$^{-1}$\ }
\begin{document}

\singlespace

\title{MULTIPLE MERGING EVENTS IN THE DOUBLE CLUSTER A3128/A3125}

\author{James A. Rose, Alejandro E. Gaba, and Wayne A. Christiansen}
\affil{Department of Physics and Astronomy, University of North Carolina, Chapel
Hill, NC 27599}
\affil{Electronic mail: jim@physics.unc.edu, gaba@physics.unc.edu, wayne@physics.unc.edu}

\author{David S. Davis}
\affil{Center for Space Research, Massachusetts Institute of Technology, Cambridge, MA 02139}
\affil{Electronic mail: dsd@space.mit.edu}

\author{Nelson Caldwell}
\affil{Smithsonian Astrophysical Observatory, 60 Garden Street, Cambridge, MA
02138}
\affil{Electronic mail: caldwell@cfa.harvard.edu}

\author{Richard W. Hunstead}
\affil{School of Physics A29, University of Sydney,NSW 2006, Australia
}
\affil{Electronic mail: rwh@physics.usyd.edu.au}

\author{Melanie Johnston-Hollitt}
\affil{Department of Physics and Mathematical Physics, University of Adelaide, 
Adelaide  SA 5005, Australia and Australia Telescope National Facility, PO Box 
76, Epping, NSW 1710, Australia}
\affil{Electronic mail: mjohnsto@physics.adelaide.edu.au}

\begin{abstract}

Multi-fiber spectroscopy has been obtained for 335 galaxies in the field of
the double cluster A3128/A3125, using the 2dF multi-fiber positioner on the
Anglo-Australian Telescope.  When combined with previously published results,
a total of 532 objects in the double cluster now have known redshifts.  
We have also obtained a 20 ks {\it Chandra} ACIS-I image of the central 16\arcm x
16\arcm of A3128 and radio imaging of the cluster with the Molonglo
Observatory Synthesis Telescope and the Australia Telescope Compact Array.
The spatial-kinematic distribution of redshifts in the field of A3128/A3125,
when combined with the {\it Chandra} ACIS-I image of 
A3128, reveals a variety of substructures present in the galaxy
distribution and in the hot intracluster medium (ICM).  The most striking
large-scale feature in the galaxy distribution is a relatively underpopulated
redshift zone $\sim$4000 \kms on either side of the mean cluster velocity at
$\sim$17500 \kms.  We attribute this depletion zone to the effect of
the extensive Horologium-Reticulum (H-R) Supercluster, within which A3128/A3125
is embedded.  In addition to this large-scale feature, numerous smaller groups
of galaxies can be identified, particularly within the underpopulated region
within $\pm$4000 \kms of the mean cluster redshift.  Due to the large
gravitational influence of the H-R Supercluster, these groups arrive at A3128
with a high infall velocity, well in excess of the local sound speed.  Two of 
these groups
appear as elongated filaments in position-velocity diagrams, indicating that
they are tidally distended groups which have been disrupted after a close
passage through A3128.  In fact, A3125 itself appears to be in such a post
passage condition.  We have identified a primary NE-SW merger axis connecting
A3128 with A3125, along which the filaments are also oriented.  In addition,
the {\it Chandra} image reveals that the X-ray emission is split into two components,
each with very small core radii, that are separated by $\sim$1 Mpc along the 
NE-SW axis.  We have combined the redshift, X-ray, and radio data to propose
that the complex X-ray morphology revealed in the {\it Chandra} image is likely the
result of a hypersonic infall of a relatively small group into A3128. The group
produces a major disruption in the ICM due to its high infall velocity.  

\end{abstract}


\section{Introduction}

The distribution of groups, clusters, and superclusters of galaxies represents a
fundamental testing ground for theories of the origin and evolution of 
structure in the universe.  Until recently, the intersection between observation
and theory has largely relied on statistical properties of
galaxy spatial/kinematic clustering, along with X-ray determined global ICM
temperatures and azimuthally averaged radial brightness and temperature 
profiles.  In contrast, much of the existing observational data
indicates that large-scale asymmetric, probably filamentary, structures,
which are not easily subjected to statistical definition {\it on an individual
basis}, are present in clusters and superclusters (e.g., Gregory \& Thompson
1978; Shandarin 1983; de Lapparent,
Geller, \& Huchra 1986; West, Jones, \& Forman 1995; West \& Blakeslee 2000).

Multi-fiber spectroscopy of galaxies with 400-fiber positioners
deployed over 2\arcdeg~ fields,
as well as the unprecedented combination of spatial resolution and sensitivity 
in X-rays provided by the {\it Chandra} and XMM-Newton Observatories, are 
rapidly advancing the observational view of clusters of galaxies.
In addition, numerical simulations, carried out within the framework of a cold
dark matter dominated universe in which structure is built up in a hierarchical
fashion, have reecently
reached the point where large areas, on scales comparable to the
largest superclusters, can be simulated in some detail (Pearce \etal 2001 and
references therein). Furthermore, using
parametrized physics of the baryonic component, simulations within the 
framework of a cold dark matter dominated universe now predict the time
evolution of the baryonic material (Kauffmann \etal 1999a,b; Somerville \&
Primack 1999; Benson \etal 2001 and references therein), thus providing a more 
direct link to the observational data.   Hence the opportunity
now presents itself to make a comprehensive comparison between observations
and simulations in the case of {\it specific} clusters/superclusters, thereby 
providing a more sensitive probe of structure evolution than can be gained 
from purely statistical considerations.  Recent studies of the Shapley
Supercluster, an extremely massive concentrations of galaxies and clusters
first pointed out by Shapley (Shapley 1930; Raychaudhury 1989),
represent one example of the possibilities provided by an
intensive campaign of multi-fiber spectroscopy, as well as of X-ray and radio 
imaging.  These
studies have produced a wealth of data on the spatial/kinematic structures
present in the Shapley Supercluster among the galaxies (e.g., Quintana \etal 
2000; Drinkwater \etal 1999; Bardelli, Zucca, \& Baldi 2001) and on the hot ICM (e.g., 
Ettori \etal 2001), as well as results on the the emission line and radio 
properties of cluster galaxies (Baldi, Bardelli, \& Zucca 2001; Venturi \etal 2001).  In
the process, a variety of structures have been found, and the influence of
environment on both galaxy and radio source evolution has been partially
elucidated.

In this paper we present new observational data on a double galaxy cluster,
A3128/A3125, which is itself embedded in the massive Horologium-Reticulum
(H-R) Supercluster.  As is discussed later, the H-R Supercluster, and 
the above-mentioned Shapley Supercluster, represent the two largest cluster
concentrations within the local 300 h$_{50}^{-1}$ Mpc\footnote{Throughout this 
paper we adopt $H_0 = 50\,h_{50}^{-1}$ \kms}.
The observational data
consist of multi-fiber spectroscopy of the double cluster A3128/A3125
with the Anglo-Australian Telescope and 2\arcdeg~field multi-fiber positioner. 
In addition, we present X-ray imaging of the
the central regions of A3128 and A3125 with the Chandra X-ray Observatory.  
Radio observations obtained with the Molonglo Observatory Synthesis Telescope
and the Australia Telescope Compact Array are also presented.
These observations have revealed a
variety of substructures, both in galaxy groupings and in the
multiple X-ray peaks detected, which in responding to the
large gravitational acceleration provided by the H-R Supercluster, reach 
unusually high infall velocities into A3128.  Thus the connection between
A3128/A3125 and the surrounding H-R 
Supercluster provides an opportunity to characterize
substructure merging on a variety of scales and unusual conditions.   
In \S 2 we present the
multi-fiber spectroscopy and X-ray and radio data.  In \S 3 we discuss the kinematic
structures revealed by the multi-fiber spectroscopy, while in \S 4 we discuss
the observed state of the ICM, as inferred from the Chandra observations, and
in \S 5 we present the MOST and ATCA radio imaging.
A discussion of our results is given in \S 6; specifically, we attempt to
produce a unified view of the merging events occurring in A3128/A3125 based on
the full set of observations.

\section{Observational Data}

\subsection{AAT/2dF Multi-Fiber Spectroscopy}

Multi-fiber spectroscopy in A3128 and A3125 (also known as DC0329-53 and 
DC0326-52 from the Dressler (1980) study) has been carried out using the 400
fiber positioner and double spectrograph system (2dF) on the Anglo-Australian
Telescope (AAT).  The field center for the observations is at 3:29:24 -52:49:45 
(J2000), so as to fully cover the A3128/A3125 double cluster system with the 
2\arcdeg \ field of 2dF.  We obtained a complete catalog of
all galaxies down to a magnitude of B$_{J}$ $<$ 18.5 in this field from the UKST
survey plates scanned by the SuperCOSMOS machine at the Royal Observatory
Edinburgh (Hambly \etal 1998, 2001).  Discrimination between stars and galaxies is
carried out by the SuperCOSMOS software, thus we restricted our sample to
those objects identified as galaxies in the SuperCOSMOS catalog.
A total of 698 galaxies to the limit B$_J$ $<$ 18.5 were found. 
However, spectroscopy of 72 galaxies has previously been obtained in the 
double cluster A3128/A3125 by
Caldwell \& Rose (1997),  and 193 galaxies have been observed in the ESO Nearby 
Abell Cluster Survey (ENACS; Katgert \etal 1996, 1998). Since 42 galaxies are 
common to Caldwell \& Rose (1997) and ENACS, a total 
of 223 velocities were available prior to our study. The ENACS velocities 
are concentrated in A3128, hence it was highly desired to increase the
number of velocities in A3125 to be able to study its kinematic
properties. 
We excluded from our SuperCOSMOS-based catalog
the 72 galaxies previously observed by Caldwell \& Rose (1997; where the
galaxies are referred to by their Dressler (1980) numbers),
as well as 22 background and 
foreground galaxies from the ENACS survey. We kept the other 171 ENACS 
galaxies, since we desired higher S/N spectra for them, but gave them a lower
priority in the fiber assignment process.  Because 42 of the 
Caldwell \& Rose (1997) galaxies are also in the ENACS list, those galaxies
were also included in the final catalog.  Altogether, the number 
of galaxies in the final catalog was reduced to 640.

The AAT/2dF system has 2 identical spectrographs each receiving 200 fibers,
for a total of 400 fibers (Lewis \etal 2000). To allocate the 400 fibers, 
including sky
fibers, we used the 2dF configuration program CONFIGURE, which 
gave a configuration at the 
telescope allowing 371 object and 24 sky fibers (5 fibers on the A 
spectrograph were unusable).   

The A3128/A3125 field was observed in service mode on 21 January, 2001 (UT). Five separate
exposures of 30 minutes were taken, using the 1200 lines mm$^{-1}$ blue
grating in first order, yielding a dispersion of 1.11 \AA/pix at the Tek5
1024 X 1024 pixel CCD, and a spectral resolution of $\sim$2.8 \AA \ (or
$\sim$200 \kms) FWHM.  The wavelength range covered is 3750 \AA \ -- 4850 \AA.
A total of 371 spectra were obtained resulting in  
358 measured redshifts; of these 23 turned out to be misclassified Galactic
stars.  Hence we obtained velocities for a total of 335 galaxies.  For 58 of
these galaxies, and for two of the stars, velocities have been published
previously by either the ENACS collaboration (Katgert \etal 1998) or 
Caldwell \& Rose (1997).
Including the other previously available velocities from Katgert \etal 1998
and Caldwell \& Rose (1997) we have in our sample redshifts for 532
galaxies in our SuperCOSMOS-derived catalog, for 76\% completeness to B$_J$ $<$ 18.5.

The data from each spectrograph were reduced independently using the 2dF 
data reduction system 2dfdr. The standard automatic reduction procedure was 
followed which automatically does the debiassing, trim-line map generation, 
fiber extraction, cosmic ray removal, wavelength calibration and sky 
subtraction. The reduced files from each exposure were then combined into
single spectra of improved S/N. 

Heliocentric
radial velocities from the absorption line spectra
were obtained for all galaxies using the fxcor routine in the
IRAF rv radial velocity package. The spectrum of a star that was 
included in the sample was used as the radial velocity
template.  The radial velocity zero-point for that star was determined by
cross-correlating its spectrum against that of several stars with known radial
velocities in the Coude Feed Spectral Library (Jones 1999), which is available
from the NOAO ftp archive.
Where possible, radial velocities were also
measured from emission line features by two of us (JAR and AEG) independently, 
by fitting a gaussian profile to all 
measurable emission lines.
An average offset of 129 \kms between emission and absorption line
velocities was found for cases where good velocities can be obtained for each.
The reason for such an offset is unknown.  However, uncertainties of 
$\sim$100 \kms in the radial velocities are small compared to the large
systematic motions found in this paper.

A final radial velocity for each object was thus obtained using the following
approach.  For cases in which the emission spectrum dominates, the final
velocity is based entirely on the emission line measurements with a correction
of +129 \kms.  In the case 
that both reliable emission and absorption velocities can be obtained (i.e.,
a standard old-galaxy spectrum is clearly visible underneath the emission), we
average the emission and absorption measurements, after first correcting the
emission measurements by 129 \kms.  In Table \ref{velocities_table} all radial
velocity measurements are listed.  In columns (1) and (2) are the reference
number from our 2dF catalog and reference numbers from previous observations
respectively.  This is followed by the RA and Dec, J2000, for each galaxy, 
the B$_J$ magnitude, and the radial velocity.  In column (7) we tabulate 
whether the velocity was based exclusively on emission lines (``e''), 
exclusively on absorption lines (``a''), or on a combination of the two 
(``ae'').  In column (8) we give the error in the radial velocity, as
given by the cross-correlation routine.  Note that this error estimate is
invalid for spectra whose velocity determinations are based on emission lines
only.  For the latter, error estimates are difficult to produce, especially
since they are very dependent on the strength of the emission lines and the
S/N ratio in the spectra.  In most cases, however, in which emission 
dominates, we have been able to locate a line with sufficient S/N ratio
that a velocity error in excess of 100 \kms is unlikely.  Finally, in column (9)
we tabulate whether any emission line is found in the spectrum.

As a test for a zero point uncertainty in our velocities we compared our 2dF
velocities with those in common with both the ENACS (Katgert \etal 1998) and
Caldwell \& Rose (1997) samples.  There are 57 galaxies for which we have 
duplicate measures with ENACS.  Omitting the two cases with large velocity
discrepancies we find a mean zero point difference of -38 \kms between the
AAT/2dF and ENACS velocities, with the 2dF velocities lower than ENACS.  The
rms dispersion between the two samples is 72 \kms.  There are also three
galaxies in common between the 2dF data and that of Caldwell \& Rose (1997), for
which there is no ENACS velocity as well.  Our mean 2dF velocity is 33 \kms 
lower than Caldwell \& Rose (1997) for those three galaxies.  There are also
42 galaxies in common between Caldwell \& Rose (1997) and ENACS.  The mean 
velocity difference is -21 \kms, in the sense that the Caldwell \& Rose (1997)
velocities are lower than ENACS.  Thus small velocity zero point errors may
exist in the data, but at a level insignificant compared to the large group
motions discussed in \S 3.

We also measured [OII]$\lambda$3727 emission line equivalent widths 
to compare emission line strengths for galaxies in different parts of
the cluster. Unfortunately, for the fainter galaxies in the sample we 
encountered considerable problems in the continuum levels in the sky-subtracted
spectra, especially in the blue end of the spectrum occupied by 
[OII]$\lambda$3727.  In many cases we find the continuum level is actually
negative, leading to an unphysical negative equivalent width for
[OII]$\lambda$3727.  We tried to remedy the situation by correlating the
SuperCOSMOS B$_J$ magnitude with the flux in the continuum of the spectra near
4500 \AA.  Our goal was to find a mean normalization factor between the B$_J$
magnitude and the continuum flux in the spectra for the brighter galaxies, and
then use that normalization factor to allow substitution of the more reliable
B$_J$ magnitude in calculating the [OII]$\lambda$3727 equivalent width from
the observed instrumental flux in the line and the blue magnitude.  However, there is such
a large scatter in the plotted relation between B$_J$ magnitude and continuum
flux in the spectra that we were unable to follow this plan.  Thus although
we are able to search to sensitive limits (typically better than 2 \AA \
FWHM) in the [OII]$\lambda$3727 line, we are unable to produce reliable
equivalent widths.  As a result, we list only whether or not emission has been
detected in column (9), but no quantitative information in regard to 
equivalent width of [OII]$\lambda$3727.  

For the same 55 galaxies in common between our sample and the ENACS sample we
find the following emission line statistics.  For 30 galaxies both ENACS and
AAT/2dF spectra show no emission lines, while for 13 galaxies both ENACS and
AAT spectra record emission.  In addition, there are 12 galaxies for which we 
find emission lines, but ENACS doesn't. There are no galaxies for which ENACS
reports emission but we do not.  Hence, our AAT/2dF spectra are finding emission
lines at approximately twice the frequency of the ENACS spectra.  To further assess
this discrepancy, for two galaxies we report emission lines, ENACS does not, and
Caldwell \& Rose (1997) also do not report emission.  Specifically, the two
galaxies are AAT\#477=ENACS\#148=CR\#80a, for which we find an equivalent
width in [OII]$\lambda$3727 of 9 \AA , and AAT\#481=ENACS\#151=CR\#53a, for
which we find an equivalent width of 6 \AA .  We have reinspected the Argus
spectra obtained in Caldwell \& Rose (1997), and now see evidence for 
[OII]$\lambda$3727 emission in both these spectra.  Both Argus spectra in
Caldwell \& Rose (1997) are clearly
of substantially lower S/N ratio in the vicinity of [OII]$\lambda$3727,
while the AAT/2dF detections are robust.  A comparison between the 2dF and Argus
spectra of the two galaxies is made in Fig. \ref{emission}, where the clear
detection of [OII]$\lambda$3727 emission in the 2dF spectra is evident.
In short, we conclude
that the better S/N ratio (especially in the blue at [OII]$\lambda$3727) of
the AAT/2dF spectra has allowed us to probe lower emission levels than in
previous analyses.

\subsection{Chandra ACIS-I Imaging}

Abell 3128 was observed on 5 May 2000 for a total of 19,513 seconds using 
the {\it Chandra} X-ray Observatory with the Advanced CCD Imaging Spectrometer
(ACIS) in its default imaging mode (ACIS-I). The field center of the 
observation is
03:30:21.93 -52:31:51.5, i.e., approximately midway between the two X-ray peaks 
previously identified on ROSAT archival images.  In ACIS-I imaging mode,
the primary field is imaged onto four front-illuminated CCD chips (referred to 
as I0-I3), arranged in a 2 x 2 array so as to cover a field of 16\arcm x 
16\arcm.  The telescope is dithered (by 16\arcs peak to peak) during the 
exposure, to provide exposure in the ($\sim$11 \arcsc) gaps between the 
CCD chips.  The results presented here are based on reprocessed
data (revision 2), and the events are filtered to include events
with flight grades 0,2,3,4 and 6 and energy $<$ 10 keV. For the imaging
analysis the lower energy range was restricted to 0.5 to 10 keV. The background
during this observation was free of flares or enhancements, so no data was
rejected because of high background. 

The imaging analysis was performed after correcting the data for
vignetting and position dependent gain effects (primarily due to charge transfer
inefficiency resulting from radiation damage) with the current
calibration files in 5 energy bands, 0.5 -- 1.5, 1.5 -- 2.5, 2.5 --
4.0, 4.0 -- 6.0, and 6.0 -- 10.0 keV. These were then combined into a
single broadband image from 0.5 to 10.0 keV. 

Background determination
for extended sources can be notoriously difficult when the source
fills most of the field of view, which is the case for this observation
(see fig 17). By examining the data in several bands and at various
smoothing scales we found two regions that are free of cluster emission, one 
on the extreme NW corner of the I0 chip and another on the far SW corner of 
the I3 chip.  These regions are used as the background
for the imaging analysis and for the spectral analysis of the cluster
emission.

\subsection{Radio Observations}

\subsubsection{MOST data}

The Molonglo Observatory Synthesis telescope (MOST) is an east-west
synthesis array comprising two colinear cylindrical paraboloids each
12~m wide by 778~m long, separated by a 15~m gap (Robertson 1991).  It
operates at a frequency of 843 MHz with a 3~MHz bandwidth.  For the
observation of A3128 the synthesized FWHM beamwidth was $43'' \,\,{\rm
(RA)} \times 54''\,\, {\rm (Dec)}$ and the field of view was $70'
\,\,{\rm (RA)} \times 88' \,\, {\rm (Dec)}$.  The data reduction used standard
in-house software and the rms noise in the CLEANed image was 0.85 mJy/beam.

\subsubsection{ATCA data}

The Australia Telescope Compact Array (ATCA) is a 6-element east-west
array with a maximum baseline of 6~km.  Simultaneous observations were
made in two frequency bands, 1384 and 2496 MHz, each of bandwidth 128
MHz.  Short observations were made with different antenna
configurations in 1999 February, 2000 March and November.  These data
sets were co-added after calibration, giving a total integration time
of 3.6 hours.


All data reduction was performed using the {\sc miriad} software
package (Sault et al.\ 1995). The primary flux density calibrator was
PKS B1934$-$638, and the phase calibrator PKS B0334$-$546.  The
resultant beamshapes were $15'' \times 10''$ in PA $-20^{\circ}$ at
1384 MHz (natural weighting), and $5.2'' \times 3.2''$ in PA
$-10^{\circ}$ at 2496 MHz (uniform weighting).


Full details of the radio observations are given in Table \ref{radio_obs}.

\section{Spatial/Kinematic Substructure in A3128/A3125}

Kinematic signatures of groups and other structures within a cluster can be
used, in principle, to assess whether clusters are indeed formed through a merger
hierarchy.  In this Section, we make an empirical examination of the
spatial-kinematic distribution of the galaxies in A3128/A3125, to look for the
presence of substructures on various scales.  In what follows we begin with
the largest scale features found in our galaxy database, and then work toward
smaller scale groups, before turning in \S 4 to substructure in the ICM.  A
key challenge will be to assess the statistical reality of apparent groupings
identified in the galaxy data.

An overall impression of the A3128/A3125 spatial-kinematic structure can be
obtained from the Dec versus RA plot in Fig. \ref{cluster}.
The RA and Dec axes are plotted in units of arcminutes from the center of
A3128.  We have plotted all galaxies from our sample
within the redshift range of 9000 $<$ cz $<$ 29000 \kms, but with those
galaxies in the narrower velocity range 16000 $<$ cz $<$ 19500 \kms plotted as
filled squares.  The locations of the two bright X-ray peaks,
discussed in \S 4, are plotted as large six-pointed stars.  There is a marked
contrast between A3128, which exhibits a clear central galaxy concentration,
and the less massive A3125, which has a dispersed appearance.  The latter
cluster is primarily located between 0\arcm and -50\arcm in $\Delta$$\alpha$ and
between -80\arcm and -25\arcm in $\Delta$$\delta$ in Fig. \ref{cluster}.
On the basis of
N-body simulations, Caldwell \& Rose (1997) argue that the dispersed nature of
A3125 is due to the tidal effect of its recent passage through A3128 (i.e.,
we are now observing A3125 {\it after} a close passage through A3128).

\subsection{Gaps in the Velocity Histogram Around A3128/A3125}

In Figs. \ref{allv} and \ref{foba_xyv} we plot position-velocity diagrams in
both RA and Dec.\footnote{For greater ease of comparison between RA and Dec 
plots we have chosen not to display the data in the form of a traditional wedge 
diagram.}  In Fig. \ref{allv} the large velocity range from 5000 \kms to
70000 \kms is included, while in Fig. \ref{foba_xyv} we emphasize the more
restricted velocity region between 9000 \kms and 29000 \kms.  The nonuniform 
distribution in velocity is readily apparent.  Aside from the obvious 
A3128/A3125 cluster peak at $\sim$18000 \kms, there are concentrations as well
at $\sim$13000 \kms, $\sim$23000 \kms, and $\sim$33000 \kms, i.e., at
-5000 \kms, +5000 \kms, and +15000 \kms with respect to the the mean cluster
velocity.  Due to our 
interest in A3128/A3125, and its relation to the surrounding H-R Supercluster,
in what follows we restrict our attention to the velocity peaks at 
$\sim$13000 \kms (or double peak at $\sim$12000 \kms and $\sim$14000 \kms), 
$\sim$18000 \kms, and $\sim$23000 \kms, which are better
seen in Fig. \ref{foba_xyv}.

To further emphasize the uneven distribution of redshifts in the direction of
the A3128/A3125 system, we plot the histogram of velocities in Fig. 
\ref{velhist}.  The most striking aspect of the histogram is the fact that 
while the velocity distribution drops off sharply, as expected, at 
$\sim$1500 \kms on both sides of the central cluster redshift at
$\sim$17775 \kms, the distribution increases again substantially at $\sim$4000
\kms on both high and low velocity sides, i.e., at $\sim$14000 \kms and
$\sim$22000 \kms.  On one hand, the dip in the histogram, followed by a subsequent
rise at both high and low velocity, is basically in accord with the
expectation that a cluster will carve out a local velocity void around it,
and that the velocity distribution then returns to the unperturbed Hubble flow 
beyond the cluster turnaround radius. On the other hand, the turnaround radius for a
cluster with velocity dispersion of 1000 \kms is expected to be at 1500 \kms 
(Kaiser 1987), while the rebound in the A3128/A3125 velocity distribution is at
4000 \kms.  As will be discussed further in \S 5, we attribute the abnormally
large gap in the Hubble flow that is inferred from the velocity histogram in
Fig. \ref{velhist} to the gravitational influence of the H-R Supercluster.
The implication is that not only is there an abnormally large void in the
Hubble flow produced by the H-R Supercluster, but that {\it groups of galaxies falling
into A3128/A3125 after detaching from the Hubble flow are accelerated to unusually
high infall velocities, due primarily to the H-R Supercluster, and secondarily
to the acceleration caused by A3128/A3125 itself.}

\subsection{Identification of Galaxy Groups and Filaments}

A second interesting aspect of the position-velocity plots shown in Fig.
\ref{foba_xyv} is the apparently non-uniform distribution of galaxies in
certain areas of the plots, which indicate the presence of spatial-kinematic
substructures in A3128/A3125.  The apparent groupings are most noticeable on
the high velocity side of the cluster.  We have identified several 
groupings, selected either on the basis of their clumping in RA-$cz$ or
Dec-$cz$ space.  These candidate groups are identified in position-velocity
diagrams in Fig. \ref{groups2_xyv}, where they are plotted with different
colors and symbols.  We have identified two types of groups; (1) those which
are fairly symmetrically distributed in position and velocity (such as the
group, with magenta open circles, centered at Dec=-60\arcm and cz=18600 \kms), and (2) those which appear
to form filaments in position-velocity space (such as the group highlighted
by red squares
which stretch from Dec=-10\arcm and cz=19300 \kms to Dec=20\arcm and cz=20300
\kms).  The locations of these groupings in RA-Dec space are plotted
in Fig. \ref{groups2_xy}, where the same colored symbols of the groups 
identified in Fig. \ref{groups2_xyv} are carried over into the RA-Dec plot.
Specifically, we have selected two candidate groups (plotted as magenta and 
green colored annuli) and two candidate filaments (plotted as red and blue 
colored filled squares).  The typical
spatial dimension of these systems is $\sim$10 - 20\arcmn, or $\sim$1 - 2 Mpc
for our assumed value of 50 km sec$^{-1}$Mpc$^{-1}$ for \hub.  The kinematic
length of the filaments, however, at $\Delta$cz$\sim$1000 \kms, implies a 
spatial length in the Hubble flow of $\sim$20 Mpc, an implication which is 
discussed in \S 6.3.

Are the groups and filaments identified in Fig. \ref{groups2_xyv}
truly physical groups or just chance associations selected by eye?
Unfortunately, testing the statistical reality of such groupings,
given that our analysis is clearly ex post facto (we first select
apparent groupings in the plots, and then statistically analyze
whether they are indeed real), is notoriously suspect.  With any
large sample of data points, it is not difficult to find a suggestive
grouping in the data, for which an ex post facto analysis produces a
high degree of confidence that the grouping is not random.  To make a
more realistic assessment of the reality of the group, one needs to
determine what is the likelihood in a data sample (of a particular
overall size) of finding at least one grouping that is as large as the
one selected.  Such an analysis in principle could be performed in the
following manner.  One could simulate a large number of clusters
similar to A3128/A3125 by randomly selecting galaxies from the
observed redshift distribution of Fig. \ref{velhist} and then
distributing them at random in RA and Dec over the region covered by
our data sample.  One could then assess how frequently groups as large
and as concentrated as the ones selected in Fig. \ref{groups2_xyv}
show up in these random simulations.  It should be clear to the
reader, however, that the distribution of galaxies in, especially, the
redshift region 18500 \kms $<$ 22000 \kms is indeed non-random and
will indicate clumping at a high statistical level.  The problem is
that by setting up a baseline of {\it complete} randomness in RA and
Dec, we are in effect setting up a statistical strawman to knock down.

We use, instead, the following statistical approach.  In principle,
the RA versus $cz$ plot is statistically independent from the Dec
versus $cz$ plot.  Therefore, if we select a candidate group or filament
that is localized in both position and velocity in the RA versus $cz$
plot, but that is actually just a false statistical group, the data
points should appear randomly spread in Dec in the corresponding Dec
versus $cz$ plot.  An example of such behavior is shown in
Figs. \ref{control_xyv} and Fig.  \ref{control_xy}, where an apparent
tight group of points localized between -30\arcm and -22\arcm in
$\alpha$ and between 17500 \kms and 18300 \kms in redshift is spread
widely in $\delta$.  In contrast, the groups and filaments selected in
either RA or Dec in Fig. \ref{groups2_xyv} are more localized in the
corresponding coordinate than the pseudo group shown in Fig.
\ref{control_xyv}.  The RA versus Dec plot of this pseudo group in Fig. 
\ref{control_xy} confirms the non-association of the galaxies.

To provide a quantitative footing to the above
approach, in each case that we find a candidate group localized in RA (or
Dec) and $cz$, we create a control sample containing all data
points within the identifed $cz$ limits for that group, excluding
the group members themselves.  We then apply three statistical
measures to assess the reality of the candidate group or filament in
three dimensional $\alpha$-$\delta$-$cz$ space.  First, we apply the
Kolmogorov-Smirnoff (K-S) two-sample test to the group versus
control sample in the following manner.  If the candidate group has been
defined in $\alpha$ and $cz$, we apply to the K-S test to the $\delta$
values, to assess the likelihood that the $\delta$ values of the
group and control samples come from the same parent
population.  If the group has been selected in $\delta$, then we
apply the K-S test to the $\alpha$ values.  Second, we make a linear
least squares analysis for a correlation between RA and $cz$, and also
test for a correlation between Dec and $cz$.  Third, we measure the
rms dispersion in both the $\alpha$ and $\delta$ coordinates for each
group, and for each control sample.  The results of the statistical
tests are given in Table \ref{groups_table}.  In column (1) the group
identification is given, where G1 and G2 are the two candidate groups (the
green and magenta colored annuli respectively); F1 and F2 are the two
candidate filaments (the red and blue colored squares respectively); C1 is
the control group from Fig. \ref{control_xyv}.  In column (2) is
given the number of group members, while in columns (3), (4), and (5)
are given the limits in $cz$, $\alpha$, and $\delta$ for each group
respectively.  In column (6) are listed the results of applying the
Kolmogorov-Smirnov two-sample test to each group versus its respective
control sample; the number given is the likelihood for the hypothesis
that the group and its control are drawn from the same parent sample.
In columns (7) and (8) are listed the results the linear least
squares regression analysis for RA versus $cz$ and Dec versus $cz$
respectively.  The numbers represent the probability that the
variables are uncorrelated.  Finally, in columns (9) and (10) are
given the rms deviations in the RA and Dec coordinates for the groups.
Immediately below each group in Table \ref{groups_table} is given the
data for its specific control sample, for which selection is made in
$cz$ only.

The two groups, G1 and G2, stand out clearly in having low $\sigma_{\alpha}$
and $\sigma_{\delta}$ compared to their control samples.  In addition, the K-S
test indicates a very high confidence that they are different from their
control samples.  On the other hand, there is no indication of correlation
in $\alpha-cz$ or $\delta-cz$,  which is expected since neither group
appears to have a filamentary elongation.  In contrast, the control group
C1, while defined to have a small spread in $\alpha$, shows a large spread
in $\delta$.  Surprisingly, the K-S test rejects at a 97.5\% likelihood the
hypothesis that the C1 group and its own control group are selected from
the same sample.  However, the C1 group was defined to have a low mean RA,
which places it away from the main concentration of A3128.  Therefore, the
distribution of Dec values in this pseudo group do not reflect that of the main
A3128 cluster.  The control group for C1 is defined in $cz$ only, hence
the distribution in Dec shows the expected concentration around A3128.  As a
consequence, the K-S test indicates that the samples are different.  Thus
although the Dec values in the C1 ``group'' are clearly spread quite randomly
(see Figs. \ref{control_xyv} and \ref{control_xy}), the K-S test does not
actually test that fact.

The case of the two filaments, F1 and F2, is less clear.  The principal 
evidence that F1 and F2 are real structures lies in the correlation 
statistics between $\alpha$ and $cz$ and between $\delta$ and $cz$.  For F1,
the probability of a correlation is high in {\it both} coordinates, while for
F2 the probability is only significant in $\delta$ versus $cz$.  A high
correlation is also found in $\alpha$ versus $cz$ for the two control samples.
This is due to the fact that the control samples for both F1 and F2 contain
parts of both filaments.  Since F1 and F2 are widely separated in $\delta$ and
also have a mean offset in $cz$, a high correlation coefficient is found.
This result reinforces the above example regarding the K-S result for the
C1 control sample, namely, that blindly applied statistical tests can be misleading.
Overall, we consider the case for the reality of F1 to be strong, given the
high correlation probability in both RA and Dec.  The statistical case for F2, 
on the other hand, rests entirely on the correlation found in $\delta$ versus
$cz$, and thus needs to be taken with caution.  We also note that G1 joins
onto F1, as does G2 onto F2.  Thus we have perhaps artificially separated
a group component from a filament component.

In addition to the G1, G2, F1, and F2 groups shown in Figs. \ref{groups2_xyv} and 
\ref{groups2_xy}, we have selected five additional small groups, each with less
than 10 members, which are plotted in Figs. \ref{groups3_xyv} and 
\ref{groups3_xy}.  For these latter groups, the only statistical test that we
have applied is to measure the dispersions in RA and Dec, as well as for their
corresponding control samples.  The results of that analysis is summarized in
Table \ref{groups3_table}.  In all cases, the small rms dispersion in one
coordinate is matched in the other coordinate, particularly when a single 
discordant galaxy is removed.  In contrast, the rms dispersion in the 
corresponding control sample is always substantially larger, except in the case
of group G6, for which the control sample of only 4 galaxies appears to 
constitute a small group of its own.

A final comment about the statistical reality of groups concerns the linear
velocity gradient in the A3125 cluster reported in Caldwell \& Rose (1997),
on the basis of 19 galaxies, which is typical of the numbers in G1, G2, F1, and
F2.  That linear velocity gradient is no longer evident in A3125, now that a
much larger sample of galaxies has been observed.  In particular, the presence
of G2 and F2 in the SW, which due to the smaller area sampled by 
Caldwell \& Rose (1997) was missed by them, has swamped the original gradient.
It appears likely that Caldwell \& Rose (1997) detected a gradient produced by
sampling primarily from two galaxy clumps with different mean redshift.  In
short, when small numbers of galaxies are involved, it is difficult to
distinguish between a real filamentary gradient and an apparent gradient
produced by the offset properties of two groups.

\subsection{Distribution of Foreground and Background Galaxies}

Turning now to the concentrations of galaxies in the foreground and
background of A3128/A3125, we have subdivided the foreground and background
into various velocity slices and plotted them in position-velocity space in
Fig. \ref{foba_xyv}, with different colors referring to the different
velocity slices, and annuli and filled squares representing foreground and
background galaxies respectively.  The positions of the color-coded galaxies are
repeated in an $\alpha-\delta$ plot in Fig. \ref{foba_xy}.  An intriguing
aspect of Figs. \ref{foba_xyv} and \ref{foba_xy} is the nonuniform
distribution of foreground versus background galaxies.  There is an overall
tendency for the foreground galaxies (colored unfilled circles) to populate the 
NE region, i.e., to concentrate around A3128, while the background galaxies 
(colored squares) lie primarily in the
southern portion of the plot.  Caution must again be exercised in making
a statistical argument about this offset between foreground and background, 
since the background galaxies, in particular, appear to be somewhat
concentrated into groups.  The existence of background group G7 in the SE has
already been pointed out in Figs. \ref {groups3_xyv} and \ref{groups3_xy}.
As a result, the foreground and background galaxy 
distributions might well result from just a few independent groups. Thus 
statistical tests for differences between the
distribution of foreground versus background galaxies, based on assuming all
galaxies to be independent points, will not accurately reflect that
situation.

To further illustrate the differences in spatial distribution between
foreground and background, we plot the galaxy redshifts versus their radial
distance from the center of A3128 in Fig. \ref{radius}.  The large
($\sim$4000 \kms) gap in $cz$ between cluster and both foreground and background
is again clearly evident.  In addition, while the foreground galaxies tend to
have radial distances in the range 0\arcm to 50\arcmn, the background galaxies
largely avoid the central 20\arcmn.

\subsection{Statistics of Emission Line Galaxies}

As previously mentioned, due to problems with the continuum levels near
[OII]$\lambda$3727 for the fainter galaxies, it was not possible to obtain
reliable emission line equivalent widths.  However, we are able to search for
[OII]$\lambda$3727 emission in the galaxies to typically 0.5 \AA \ levels in
equivalent width.  As a result we have compiled statistics of the percentage
of galaxies with observed emission, as a function of their environment.  Here
we have restricted the discussion to only those galaxies with AAT/2dF spectra,
since those spectra provide the most homogeneous set to work from.  

We find a surprisingly high percentage of galaxies in the main A3128
cluster have detectable [OII]$\lambda$3727 emission.  Specifically, 
41 of 85 galaxies, or 48\%, with radial velocity between 16000 \kms and
19200 \kms, and with Declination North of -53:01:30 (the center of A3128 is at
$\delta$ = -52:31:30, J2000)
have emission.  This is an unusually high level of emission, given that most
surveys of emission in rich clusters report emission detection rates below
15\% (Biviano \etal 1997).  In contrast, Biviano \etal (1997) report emission
line detections for only 30 out of 152 galaxies in their data on A3128, i.e.,
an $\sim$20\% detection rate.  Thus the high fraction of galaxies found with
emission lines is partly the result of the high S/N ratio achieved for
the 2dF spectra, but also partly due to the fact that A3128 has an unusually
high level of emission, since the 20\% detection rate reported by 
Biviano \etal (1997) is at the high end for the ENACS cluster sample.  

While the main cluster emission-line galaxy fraction is high, the detection 
rate in our 2dF data of galaxies in
foreground and background flows is substantially higher.  A total of 38 out of
54 galaxies, i.e., 70\%, of the galaxies in the background within the velocity 
range 22100 \kms to 28500 \kms have detected [OII]$\lambda$3727 emission, while
15 out of 16, or 94\%, of the foreground galaxies in the velocity range
10000 \kms to 15000 \kms are also detected.  The statistics of the group and
filament galaxies is that 32 out of 48 of them,i.e., 67\%, have detected
emission.  Thus, while the emission line statistics of the main A3128 cluster
are remarkably high, indicating perhaps an overall dynamical youth of the 
cluster, the emission levels are still lower than in the immediate field and
in infalling groups.  Finally, of the 19 galaxies in the background redshift
peak at cz$\sim$33000 \kms, 16 of them (i.e., 84\%) have detected 
[OII]$\lambda$3727 emission, hence indicative of a field population.
Note that the detection of [OII]$\lambda$3727 emission alone does not allow
us to distinguish between emission due to star formation and that caused by an
active galactic nucleus.

To conclude at this point, data on the distribution of
galaxies in the field of A3128/A3125 indicates that spatial/kinematic
structures are present on a variety of scales.  The most evident structure
is the $\sim$4000 \kms underpopulated velocity region on both sides of the
cluster velocity.  Much smaller structures are evident as well, ranging from
groups of 4 galaxies to groups with a couple of dozen members, and even the
foreground and background galaxies at 4000 \kms on either side of the A3128
systemic velocity are nonuniformly distributed.  The
majority of galaxies in the underpopulated velocity regions appear to be 
members of groups and/or filaments.  However, the statistical reality of these 
groups (and particularly of the filaments) is difficult to establish. 
Most of the groups identified have high velocities with respect to the 
A3128/A3125 double cluster.  Such high velocities of these infalling and
merging substructures should in principle lead to unusual transient physical 
conditions in the hot intracluster medium, a subject which we turn to in the
next section, where we consider the Chandra X-ray images of A3128.

\section{X-ray Substructure in the Intracluster Medium}

Aside from spatial-kinematic substructures evident in the galaxy distribution,
substructure in the hot ICM also provides clues to past and ongoing cluster
merger events.  The ICM has the particular advantage of not suffering from the
discreteness problem that afflicts the statistical selection of small galaxy 
groups.  Consequently, we now turn to the {\it Chandra} ACIS-I observations
of the central region of A3128.

The X-ray morphology of the central 16\arcm x 16\arcm of A3128 is
illustrated in Figs. \ref{ds9_3} and \ref{ds9_2}, where
the 20 ks Chandra image is displayed at two different contrast
levels, for photons in the energy range 0.5 keV to 10 keV.  In all
cases the original image has been 4 x 4 binned into 2\arcs x 2\arcs
pixels, and then further smoothed with a gaussian of $\sigma$=4
pixels.  The most evident feature of the cluster X-ray morphology is
the presence of two bright X-ray peaks, aligned in the NE-SW axis, and
separated by 12\arcm, or 1.2 h$_{50}^{-1}$Mpc. 
These two peaks have been marked previously on figures such as
Fig. \ref{cluster}.  We hereafter refer to these two components as the
NE and SW peaks.  Also evident is a lower surface brightness
component, located between the two bright peaks. This third component
to the X-ray flux has a center that, to within the uncertainties, is
indistinguishable from the center of the galaxy distribution in A3128.
To further assess the morphology of the X-ray emission, we present in
Fig.  \ref{adapt_1} an adaptively smoothed version of the Chandra image.
Specifically, a maximum smoothing of $\sigma$=5
pixels was applied.  
Furthermore, to illustrate the connection between the X-ray emission and
the galaxy distribution, we have plotted the Chandra X-ray contours on
top of a grayscale representation of the optical B$_J$ image
in Fig. \ref{xrscos}.  It is evident that while the SW X-ray peak is
coincident with a bright galaxy, and with an apparent compact galaxy group,
there is no such general coincidence between the NE X-ray peak and the
galaxy distribution, a point that we return to in \S 6.

A second noteworthy feature of the X-ray morphology is the fact that
the two bright peaks have small core radii and are offset from the
lower surface brightness emission on which they are superimposed, as
can be seen qualitatively in, e.g., Fig. \ref{adapt_1}.  To quantify
the small core radii we have plotted in Fig. \ref{profs} the {\it
azimuthally averaged} radial surface brightness profiles of the NE and
SW X-ray components, plotted both in terms of linear radial distance
and logarithmic radial distance.  Also plotted are the best $\beta$
model fits to the NE and SW components, following the prescription of
Cavaliere \& Fusco-Fumiano (1978), i.e., we fit the surface brightness
distribution according to

    $\Sigma(R) = \Sigma(0)[1 + (R/a)^2]^{-3\beta+1/2},$

\noindent where $a$ is the core radius of the fit.  We obtain a core
radius of 26 h$_{50}^{-1}$kpc and 40 h$_{50}^{-1}$kpc for the NE and SW
components respectively.  As will be further discussed in \S 5, these
values are at the low end of the large range in core radii observed in
clusters of galaxies (e.g., Jones \& Forman 1984; Peres \etal 1998;
Mohr, Mathiesen, \& Evrard 1999), where a typical cluster has a core
radius of $\sim$250 kpc.  In addition, the $\beta$ values of $\sim$0.3
derived from the fits are low compared to typical values for clusters
of $\sim$0.7.

To better illustrate the fact that the bright NE and SW cores are offset from
their underlying lower surface brightness emission, we have fit elliptical 
isophotes to the NE and SW
components at various surface brightness levels, using the `ellipse' routine
in the STSDAS `isophote' package.  The fits return the X-Y positions of the
center of the fitted ellipse at each contour level, and the results have been 
plotted in Fig. \ref{centers}.  The X (RA) and Y (Dec) centers change by 55
pixels (110\arcs) and 20 pixels (40\arcs) respectively in the NE peak as one
moves from the central peak out to a maximum of 200\arcs in the semi-major axis
of the azimuthally averaged radial profile.  In the SW peak the X and Y offsets
are 25 pixels and 12 pixels respectively.  Thus the effect is considerably more
pronounced in the NE component, but clearly is present in the SW component as
well.  Moreover, while the shift in isophote center is fairly linear with
semi-major axis, in the case of the NE component, the shift is very small
within the central 20 pixels (40\arcs), then ramps up at larger radii.  Note 
that the displacement 
of the core emission in both peaks relative to the lower surface brightness
emission underlying them is in the sense that the emission falls off sharply
on the side pointing away from the cluster and trails off more slowly in the
cluster center direction.

A final morphological feature of interest is the resolved inner
structure of the NE component, seen in Figs. \ref{ds9_3} and \ref{adapt_1}.
The most striking feature of the NE
emission peak is its strong elongation {\it along} the axis connecting the NE 
and SW components.  This elongation factor is substantial, as can be seen from
the ellipticity and position angle profiles plotted in Fig. \ref{ellpa}.  While the mean 
ellipticity for the fainter contours is $\sim$0.4, it rises to 0.65 in the
inner 5 pixels (17 h$_{50}^{-1}$ kpc) in radius.  Furthermore, there is a substantial
twist in the position angle of the fitted ellipse, from 60$\arcdeg$ E of N in the
inner 10\arcs to 70$\arcdeg$ E of N at a semi-major axis of 50\arcs.  In short,
transient morphological features are present in the X-ray emission from
the ICM of A3128 on a variety of scales, from less than 20 h$_{50}^{-1}$ kpc in the
core of the NE component to the $\sim$1 Mpc projected separation between the NE and SW
components.

Another key diagnostic of the physical conditions in the ICM of A3128 is the 
temperature structure of the X-ray emitting gas.  We have derived
temperatures from the observed X-ray emission by using XSPEC (Arnaud 1996) to
fit a MekaL (Mewe \etal 1985, 1986; Liedahl, Osterhedl, \& Goldstein 1995) thin
plasma model with foreground absorption to the spectral data.  The abundances
are given relative to the solar values (Anders \& Grevesse 1989).  In these
fits the HI absorption, temperature, and abundance are allowed to vary, and we
restrict our analysis to the 0.5 -- 10 keV band.  The NE and SW
components have global temperatures of 3.9 keV and 3.6 keV respectively, while
the third, lower surface brightness, component has a temperature of $\sim$4 keV.
Moreover, while the SW component appears to be isothermal, the data on the NE
component suggests that the temperature is higher in the inner 45\arcs in 
radius.  The data on temperature and metal abundance, along with their 90\%
confidence limits, are summarized in Table \ref{temps_table}.  While the
uncertainties are substantial, the data are suggestive of a 
temperature increase in the core of the NE component, and are in any case
inconsistent with a substantial temperature drop in either the NE or SW
component.  Thus there is no indication of a cooling flow in either peak.

From the $\beta$ model fit to the radial surface brightness in the NE and SW
components, and assuming the temperatures from the ``global'' fits to each
component, we have calculated an estimated ``virial'' mass enclosed within
a radius of 0.3 h$_{50}^{-1}$ Mpc for both components, by assuming
hydrostatic equilibrium.  For the NE and SW components we then derive a
virial mass of 1.6 x 10$^{14}$ M$_{\sun}$ and 1.5 x 10$^{14}$ 
M$_{\sun}$, respectively.  On the other hand, the gas mass in the NE and SW
components (inside of 3\arcmin \ or 0.3 h$_{50}^{-1}$ Mpc in radius) derived from the
central electron densities of 1.1 x 10$^{-2}$ cm$^{-3}$ and 2.4 x 10$^{-2}$ 
cm$^{-3}$
is 6.5 x 10$^{12}$ M$_{\sun}$ and 7.2 x 10$^{12}$ M$_{\sun}$, respectively, for 
the NE and SW components (the X-ray luminosities of the two components are 
5.4 x 10$^{43}$ erg/sec and 3.4 x 10$^{43}$ erg/sec in the 0.5 -- 10 keV 
band).  Thus the implied gas-mass to
virial-mass ratios are $\sim$4-5\% in the two components, which is very low
compared to more typical values of $\sim$20-30\% usually found in rich clusters
(e.g., Mohr, Mathiesen \& Evrard 1999).  Seen another way, for the observed
X-ray emitting gas masses, the virialized velocity dispersion in
the galaxies for the NE and SW components 
should be 437 \kms and 425 \kms respectively.  While the observed
velocity dispersion of the galaxies cannot be readily separated into NE and SW 
components, and while in fact the presence of so much substructure in A3128
makes it difficult to define a global cluster velocity dispersion, it is
clear in any case that the velocity dispersion of the galaxies in the center
of A3128 is far in excess of 430 \kms.  Thus the discrepant values extracted
from a dynamical analysis indicates that the X-ray emitting gas is {\it not}
virialized and therefore must be well removed from hydrostatic equilibrium.

The third, diffuse emission component, seen between the two X-ray peaks 
(Fig. \ref{ds9_2} and Fig. \ref{adapt_1})
is noteworthy in that its extent is approximately that of a normal
cluster ($\sim$3 Mpc), it is hot (kT$\sim$4 keV), and is also enriched with an
abundance of $\sim$ 0.6 solar. The fact that the diffuse gas is
enriched implies that it is or was bound to a system with stars which 
can supply the metals. Since there are two peaks with similar
properties, determining which one might be associated with the diffuse
emission component must be based on indirect evidence. From the global
abundances in Table \ref{temps_table} we see that the NE component has a very low
global abundance, $\sim$0.13 solar, much lower than the best fit
abundance of the diffuse component, which on the other hand is consistent with
the abundance of the SW clump. So, while we cannot rule out that the
diffuse component is associated with the NE component, it seems more likely
to be associated with the enriched SW component.  Thus the third, diffuse,
component appears unlikely to have a common origin with the NE component.

Finally, we have obtained a direct estimate of the redshift of the X-ray gas
in the NE, SW, and diffuse components using the X-ray emission lines present in
the ACIS-I spectra.  For the SW component, our best fit model, in which we allow
the redshift to be a free parameter, produces a best fit redshift of z=0.065,
with 90\% confidence limits between 0.050 and 0.078.  For the diffuse emission
component, the best fit redshift is z=0.059, with a 90\% confidence range of
0.047 to 0.072.  Due to the lower counts and lower metal abundance in the NE
component, the redshift of that gas is less well constrained.  The best fit
redshift is 0.10, with a 90\% confidence range of 0.037 to 0.23.  Thus we can
conclude that the gas in the SW and diffuse components are unambiguously
associated with A3128, while the redshift found for the NE component is 
consistent with A3128, but the association is not conclusive. 

\section{Radio Sources in A3128}

Further insight into the events transpiring in A3128/A3125 can be
gained from examination of radio images of the double cluster, since
radio sources can be reenergized by ICM shocks produced in mergers
(Ensslin \etal 1998; Roettiger, Burns, \& Stone 1999; Govoni \etal
2001).  Furthermore, the morphology of tailed sources can be useful in
assessing the relative direction of motion of the source with respect
to any gas dynamical movement of the local ICM.  We begin by
displaying in Fig.\ \ref{xrmost} the MOST 843 MHz radio contours
superposed on the Chandra ACIS-I X-ray image.  Four of the radio
sources have X-ray counterparts, three of which appear to be cluster
members, as revealed by the overlaid radio-optical plot in Fig.\
\ref{mostscos}.

We will focus here on the two radio sources aligned with the NE and
SW X-ray peaks.  The weak unresolved source coincident with the SW
X-ray core is identified with ENACS\#75, one of the brightest galaxies
in the cluster.  On the other hand, the radio source at the NE peak is
both stronger and more remarkable.  It appears to consist of a bright
point source and a much fainter arc-like feature, displaced to the NE of the
bright source.  At the higher resolution of the ATCA 20 cm and 13 cm
images, however, the bright radio source resolves into an elongated
north-south double, as shown in Fig. \ref{atca_20_13}, where the ATCA
20-cm and 13-cm contours are superposed on a grayscale rendition of
the CTIO 0.9-m R bandpass image.  This figure shows that the radio
source is identified with a faint galaxy that is undoubtedly
background to A3128.  The spatial alignment of optical galaxy, MOST
and ATCA radio contours, and X-ray contours is illustrated in Fig.\
\ref{overlay_ne}
CTIO 0.9-m B and R images of the region in Fig.\ \ref{negal} show that
the faint galaxy has a concentrated red nucleus.  In the deeper R-band
image a somewhat irregular low surface brightness fuzz surrounds the
bright nucleus, 
while in the B image, a long curved structure is seen primarily to the
southwest.  A hint of this curved structure is seen in the R image as well, but
it clearly has a very blue color relative to the nucleus of the galaxy.
In the unlikely circumstance that the galaxy is a member of A3128, it
would be a low-luminosity dwarf, and hence extremely unlikely to
harbor a compact luminous double radio source.  Thus we infer that the
bright double is in fact a superposed background source that has no
physical relation to A3128\footnote{If we assume a redshift $z\sim
0.3$, consistent with the galaxy's R magnitude (see Section 6.4.6),
the integrated radio power at 1.4 GHz is $\sim 2\times
10^{25}$~W~Hz$^{-1}$, making it a powerful source, near the boundary
between Fanaroff-Riley classes FRI and FRII}.  On the other hand, the
faint extended radio arc seen in the MOST image (but not in the higher
resolution ATCA 20 cm image), is more intriguing. It is situated
slightly to the NE of the core emission in the NE X-ray component,
about where one might expect a bow shock to be present if the NE
component were plowing supersonically through A3128.  The fact that
the radio arc is displaced from the bright background source suggests
that it truly is a structure associated with A3128, but deeper radio
imaging is needed to better assess that conjecture.

\section{Discussion}

The single most compelling feature of the combined 2dF spectroscopy and Chandra
X-ray imaging is the wide variety of structures observed in both the 
spatial/kinematic distribution of the galaxies and the hot ICM.  On the
largest scales there are peaks in the velocity distribution, separated
by $\sim$4000 \kms, i.e., over inferred distances of $\sim$80~h$_{50}^{-1}$~Mpc,
while on the smallest scales the X-ray image of A3128
shows transient structures $\lesssim$20~h$_{50}^{-1}$~kpc in size.  In this section 
we attempt to draw a coherent picture encompassing the structures found on all
scales, and to establish a link between galaxy substructure and the disturbed
state of the ICM.  In what follows, we begin with the largest structures, to 
establish the overall potential field in which the smaller substructures are 
embedded, and then work toward the smaller scales.

\subsection{4000 \kms Velocity Gaps and the Horologium-Reticulum Supercluster}

As has been pointed out before, the velocity histogram in Fig. \ref{velhist}
indicates that on either side of the huge velocity peak at $\sim$17750 \kms,
representing the velocity center of A3128/A3125, there is a large ``gap'' of
$\sim$4000 \kms, in which the galaxy density is low, picking up again at
$\sim$14000 \kms on the low velocity side, and at $\sim$22000 \kms on the high
velocity side.  It is natural to assume that the increase in galaxy density
at $\sim$14000 \kms and $\sim$22000 \kms represents the resumption of the
Hubble flow on either side of the double cluster.  While this assumption might
be compatible with the galaxy distribution in redshift on the low velocity side,
the distribution at high redshift is puzzling.  There is a strong peak in
redshift at $\sim$23000 \kms, but then the histogram falls off precipitously to
higher redshift, and between $\sim$25000 \kms and $\sim$32000 there are very
few galaxies found.  The histogram then rises to another peak between 
$\sim$32000 and $\sim$35000 \kms, before dropping off once again to higher
redshift.  We note that the spatial distribution of galaxies in the redshift
peak at $\sim$33000 \kms show no indication of being concentrated into a
background cluster.  Moreover, as mentioned before the fraction of emission line
galaxies in the $\sim$33000 \kms peak is 84\%, indicative of a field, rather 
than a cluster, population.  The behavior of the redshift histogram over the 
full range out to cz = 65000\kms is shown in Fig. \ref{hist200_all}.

The velocity peaks at $\sim$14000 \kms and $\sim$23000 \kms are quite
narrow, certainly less than 5000 \kms in full width, indicating that they
might represent galaxy clusters at these redshifts. However, the galaxies in the
peaks are distributed over the full 2\arcdeg \ field of the 2dF observations,
and show no concentration into clusters.  Rather, as was discussed in \S 3,
there is some evidence to indicate an overall NE-SW asymmetry between the
distribution of low-velocity peak versus high-velocity peak galaxies,
which may result from the clumping of these galaxies into groups.  Thus, while
it is perhaps problematic to associate the peaks at $\sim$14000 \kms and 
$\sim$23000 \kms with the resumption of normal Hubble flow outside of
A3128/A3125, {\it we conclude here that the low and high velocity peaks on either
side of the double cluster are the working ``surfaces'' from which new
galaxies are being recruited into the A3128/A3125 system.}   The large,
$\sim$4000 \kms, velocity gap between these surfaces and the central cluster
velocity is far in excess of that expected from a cluster of normal mass, and
can only be attributed to a much greater massive concentration (Kaiser 1987).  As has been
pointed out previously, A3128/A3125 is embedded in the 
Horolgium-Reticulum Supercluster, which extends over an area at least 7\arcdeg \
in extent (Lucey \etal 1983).  The H-R Supercluster and Shapley Supercluster
are the two largest known mass concentrations in the local universe, i.e.,
out to 300 h$_{50}^{-1}~$Mpc (Zucca \etal 1993; Einasto \etal 1994; Hudson \etal 
1999).  In fact, Zucca \etal (1993) and Einasto \etal (1994) consider the H-R 
Supercluster to encompass 18 and  32 clusters respectively.  Thus we associate 
the largest structures in our data, i.e., the
$\sim$4000 \kms gaps to the low and high velocity of A3128/A3125, with
the huge gravitational potential well of the H-R Supercluster.  The most
important implication of this conclusion is that groups of galaxies which
are detached fom the working ``surfaces'' are accelerated to unusually
high infall velocity on their way toward  A3128/A3125, due to the acceleration
produced by the deep potential well of the H-R Supercluster.  
Specifically, the infall velocities will be 
supersonic with respect to the local sound speed in the A3128/A3125 complex,
which, at an electron temperature of 4 keV, corresponds to $\sim$1000 \kms.

To put the redshift histogram of the A3128/A3125 region into context, there
are strong similarities with the observed distribution in redshift of galaxies
in the Shapley Supercluster found by, e.g., Quintana \etal (2000) and
Drinkwater \etal (1999).  In the case of the Shapley Supercluster, the area
surveyed is very extensive, so that the complexity of the large gaps in
redshift, similar in size to the 4000 \kms gaps found by us in the H-R
Supercluster, is readily seen in, e.g., Figs. 6 and 7 of Quintana \etal (2000).
In contrast, where a large body of redshift data has been collected in the
vicinity of a cluster that is {\it not} associated with a mass concentration on 
the scale of the H-R Supercluster, such as the Coma cluster (Geller, 
Diaferio, \& Kurtz 1999), no large
gaps are evident in the distribution of velocities.  Thus the
association of the velocity gaps in A3128/A3125 with the potential well of
the H-R Supercluster is further supported by comparison to other cases.

\subsection{The A3128/A3125 Double Cluster System}

The next clearly defined scale below that of the H-R Supercluster 
is the double nature of the A3128/A3125 cluster system.  Much of this
has been discussed in Caldwell \& Rose (1997), hence we repeat only a few key
points.  The projected separation between the center of A3128 and the very poorly-defined
center of A3125 is on the order of 1\arcdeg, which corresponds to $\sim$6~Mpc.
Surprisingly, A3125 and A3128 have nearly the same mean cluster velocity, 
i.e., there is little relative motion between them along the line of sight.
Thus one cannot reconcile the large distance between the two clusters
with the small velocity separation (without appealing to a highly preferred
viewing angle), {\it unless A3125 is currently near turnaround after its passage
by A3128.}
It is difficult to define reliable global velocity dispersions for 
A3125 and for A3128, given the large number of groups and filaments present
whose inclusion or exclusion greatly affects the derived $\sigma$ (and given the
fact that A3125 is not a clearly defined cluster).  However, it is
certainly clear that the velocity dispersion of A3125 is lower than that of
A3128, which is in accord with the fact that A3128 contains $\sim$4 times
more members than A3125.  In addition, while A3128 has a well-defined central
concentration in the galaxy distribution, A3125 presents a dispersed appearance.
Based on N-body simulations, Caldwell \& Rose (1997) attribute the scattered
appearance of A3125 to a tidal passage of the latter past A3128.
In addition, while the 20~ks Chandra exposure of A3128 reveals extensive
emission from a hot ICM, the 10~ks exposure of A3125 reveals no extended
emission.  What appears to be extended emission in an archival ROSAT 4~ks
image is actually the merging of badly degraded point sources, due to the
large off-axis location of A3125 in the FOV of the ROSAT data.  If there were 
any structures
in A3125 as bright as the NE and SW components seen in A3128, they would
easily be detected in the 10~ks Chandra image of A3125.  Consequently,
the available evidence is consistent with A3125 being a relatively small cluster
that has experienced a damaging encounter with A3128.

\subsection{Intermediate-Sized Groups and Infall into A3128}

We turn now to the intermediate-sized groups (10-30 observed members) and
filaments.  Naturally, such groups are very difficult to locate in the crowded
regions of A3128 and A3125, if they do exist there, but are readily seen in
the low density 4000 \kms `gaps' between the clusters and the higher density
regions outside the H-R Supercluster.  While identification of galaxies with
groups and filaments can be problematic on statistical grounds, especially in
regard to filaments, nearly all galaxies present in the
high-velocity `gap' region can be identified with intermediate-sized and small
groups or filaments.  The two filaments are particularly interesting.  While
their spatial extent of $\sim$20\arcm corresponds to a length of 2~h$_{50}^{-1}$~Mpc,
their velocity distension of $\sim$1000~\kms corresponds to 20~h$_{50}^{-1}$~Mpc in
terms of Hubble flow.  Since it is highly unlikely that the filaments would be
so nicely oriented along our line of sight with A3128/A3125, we conclude that
{\it their distension in velocity is unlikely to be a reflection of Hubble flow,
but rather is due to a tidally-induced velocity gradient in an infalling group.}
Furthermore, it was demonstrated in Caldwell \& Rose (1997) that tidal 
distension of subclusters occurs only after the subcluster has passed through
the main cluster, in either a head-on, or slightly off-center, encounter.  
Consequently, we expect that the tidally
distended filements have already made a passage through the H-R Supercluster
and A3128/A3125 system and are now emerging from that passage.  In contrast,
the more circularly symmetric (in both position and velocity) groups are 
probably still in the infall phase of their passage through the 
cluster/supercluster.  Consequently, the filaments are more likely to be
associated with visible signs of transient merging activity in the A3128/A3125
system, especially in regard to disturbances in the ICM.  To further assess
this hypothesis regarding groups versus filaments,
it would be of interest to compare the physical properties of the galaxies
in groups versus filaments for indications of different recent evolutionary
histories.

\subsection{Small-Scale Structure in the Hot ICM}

Over length scales $\lesssim$100~kpc, and group membership
less than a few galaxies, galaxies are of limited use as tracers 
of substructure in A3128/A3125.  At this point, X-ray emission from the ICM
becomes our primary means to delineate small-scale structure. As has been 
discussed in \S3, ample evidence is seen in the Chandra image of A3128 for 
structures in the ICM on scales below 1~h$_{50}^{-1}$~Mpc.  We begin by briefly
recapitulating those key observations related to the structure of the ICM,
namely:

\noindent (1) The X-ray emission shows two distinct regions with high surface
brightness cores, and an additional lower surface brightness third component
between the two main components.

\noindent (2) The two bright components have small core radii.

\noindent (3) The bright peaks, especially the NE one, are displaced with 
respect to their underlying lower brightness contours.

\noindent (4) In addition to the overall displacement of brighter versus
fainter contours, the NE X-ray component shows a strongly asymmetric morphology.

\noindent (5) The heavy element abundance of the gas in the NE X-ray component
($\sim$0.13 solar) is significantly lower than for the gas in the SW component
($\sim$0.6).

\noindent (6) A bright galaxy (and member of A3128) is coincident with the SW 
X-ray peak; no bright galaxy coincides with the NE peak.

\noindent (7) Radio sources are coincident with both X-ray peaks.

We now attempt to produce a coherent picture of the X-ray structures, and their
relation to the optical galaxy and radio data.  We will first argue that the
X-ray structures are transient, and likely associated with merger activity.
Then we propose that the NE X-ray peak is the leading edge of highly
supersonic gas, once associated with the F1 and G1 galaxy filament/group,
which has now passed through the center of A3128.  Finally, we argue that
the SW X-ray peak is most likely a gaseous component at the trailing end of 
the F1/G1 infall pattern which is still at least partially bound to the 
underlying potential well associated with a compact group of galaxies.

\subsubsection{Transient Nature of the ICM}

The most conspicuous
aspect of the ICM is the division of the X-ray emission into two
bright peaks, separated by 1.2~h$_{50}^{-1}$~Mpc along roughly the same NE-SW axis
that joins A3128 with A3125.  A third fainter and less centrally concentrated
component is located between the NE and SW bright peaks, and very close to the
center of the galaxy distribution.  The existence of multiple X-ray components
on relatively small spatial scales implies transient disturbances in the ICM.
In addition, the two bright components have core
radii of $\sim$30~h$_{50}^{-1}$~kpc.  While most clusters typically have core
radii 10 times greater than this, there are two kinds of clusters with small
core radii.  First are the cooling flow clusters, which are generally 
characterized by a cool peak of bright emission atop a relaxed ICM with large
core radius (Mohr, Mathiesen, \& Evrard 1999).  An example of such a cooling
flow cluster with highly structured X-ray emission and a small core radius
is A1795 (Fabian \etal 2000).  Since the central temperature
does not drop in either the NE or SW peak, nor is the underlying cluster relaxed
(given the double/triple nature of the X-ray emission and the extensive
evidence for ongoing merging in A3128/A3125), the identification of the bright
X-ray components in A3128 with cooling flows is clearly inappropriate.  The
second type of cluster with small core radii is represented by those, such as 
A548, which are undergoing merging events (Davis \etal 1995).  The
similarity between A3128 and A548 is striking, in that A548 also has multiple
X-ray emission peaks, with small core radii. In this case the small core 
radii and low value of $\beta$ is considered a result of the truncation of the 
cluster profile (Mulchaey 2000; Navarro, Frenk, \& White 1996; Bartelmann \&
Steinmetz 1996).  Finally, 
given that the sound travel time across the asymmetric 20~kpc NE region
for a 4 keV gas is only $\sim$2 x 10$^7$ years, then unless the gas in the 
NE emission peak is pressure confined, we are witnessing a highly transient 
structure in the ICM.  We argue in the Appendix A.2 that the group gas has
been stripped from the group and thus is no longer gravitationally bound, but
is presently ram pressure confined.  Furthermore, the NE
component is elongated by a $\sim$2:1 axis ratio along the same NE-SW axis
that separates the two bright components.  Thus we observe a preferred
alignment along the NE-SW from scales below 20~h$_{50}^{-1}$~kpc to the
6~Mpc distance separating A3128 from A3125; alignments over a large variety of
scales is seen in other clusters as well (West \& Blakeslee 2000).  These
large-scale alignments are believed to represent the preferred
axis of infall and merging activity in the cluster.

\subsubsection{Connection Between X-ray Structure and Merging Galaxy Groups}

Having established the transient nature of the X-ray structures and its
probable relation to merging events, we now attempt
to link the major X-ray structures with specific spatial/kinematic substructures
in the galaxies.  First, it is natural to associate the double X-ray structure with
the A3128/A3125 double system.  However, the projected separation between A3128 and
A3125, at $\sim$6~h$_{50}^{-1}$~Mpc, is approximately 5 times larger than the
observed separation between the two high surface brightness X-ray peaks.  
Furthermore, the small velocity difference between A3125 and A3128 implies a
multi-Gyr timescale for that passage, whereas the timescales for
the two X-ray peaks (e.g., the cooling timescale, which is only $\sim$1 Gyr) 
appear incompatible.  Hence we look to smaller substructures in the galaxies 
for an explanation of the X-ray peaks.

In examining the position-position and position-$cz$ 
plots in Figs. \ref{groups2_xy} and \ref{groups2_xyv}, the two substructures
that are at least projected close to the positions of the bright X-ray peaks
are the group denoted by green annuli (G1 in Table \ref{groups_table}) and the
filament denoted by the red squares (F1 in Table \ref{groups_table}).  
If F1 has been correctly identified as a post-passage
(through A3128) tidally distended group, then it represents the most natural
candidate for producing the transient structure seen in the Chandra image
of the ICM.  Referring to Fig. \ref{groups2_xy} it is evident that the projected
axes of F1 and the NE-SW X-ray sources are approximately parallel.  Moreover,
since the distinction between F1 and G1 has been drawn somewhat artificially,
we henceforth refer to them collectively as the F1/G1 group.
A closer look at Fig. \ref{groups2_xy} reveals that the F1/G1 group extends
$\sim$30\arcm to the NE of the NE X-ray peak, and that this represents the
high velocity, hence likely leading, edge of the group.  Given our
interpretation of the filaments as post-passage tidally distended groups, 
we argue that the NE end of F1 has passed through the center of A3128, and that 
much of the group gas is continuing through, with some of it trailing behind, as
indicated by the elongated isophotes.
On the other hand, the bulk of G1 and the gas associated with the SW X-ray 
peak are located on the other side of the cluster center, 
and thus are considered to be still infalling.  

\subsubsection{Analysis of the NE X-ray Component}

Support for the above scenario
comes from the morphology of the NE peak -- in particular, its displacement
by more than 2\arcmn, or 200~h$_{50}^{-1}$~kpc, to the NE of the lower surface 
brightness isophotes of that X-ray component.
Given that the large acceleration imparted by the H-R Supercluster should
result in supersonic impacts of infalling groups into the main cluster, we
can expect to observe signs of shock structures in the ICM.  The large
offset nature of the NE peak may well represent such activity, in
which case the outward direction of its motion from the cluster center has
been established, in agreement with our hypothesis about the outward flow of
the NE end of F1.  Further, from the projected separation between the NE end of F1 and the
outer edge of the NE X-ray core we can derive an estimate of the relative 
velocities of the F1 filament and the X-ray source.  This calculation is
discussed in Appendix A.1.  Here we simply report the main result, which is that
the gas in the NE X-ray peak is inferred to be travelling at about half the
velocity of the leading edge of F1, and its velocity with respect to the
A3128 cluster is $\sim$3500 \kms.  This speed corresponds to a Mach Number, M=6,
for a typical cluster gas temperature of 3 keV.  Thus, as discussed in
Appendix A.2, the gas in the NE peak
is inferred to be highly supersonic, shocked gas which has been stripped from
a group of galaxies (now seen as F1) during a passage through A3128
at hypersonic velocity (in excess of 4000 \kms).  As is discussed in Appendix
A.3, at the high Mach Number (M=6)
inferred, the X-ray structure should be primarily determined by the effects of
ram pressure (which is of course axisymmetric) as opposed to cluster thermal
pressure (which is spherically symmetric).  This is clearly consistent with
the observed axial structure in the X-ray source.

One further piece of observational support for this scenario comes from the
aforementioned radio arc, which lies slightly to the NE of the peak emission
in the NE core, i.e., where a bowshock is expected if the gas is moving
hypersonically.  As mentioned previously, however, deeper
radio imaging is required to verify whether or not the radio arc is indeed 
related to the cluster, rather than to a background source.

\subsubsection{Analysis of the SW X-ray Component}

The most intriguing aspect of the X-ray structure is the existence of
two bright peaks.  The question that naturally arises is whether both
peaks can be attributed to a single merging event or whether two
separate events must be invoked.  The answer to this question is of
great significance to interpreting the events taking place in A3128.
On the one hand, our analysis has uncovered evidence for a variety of
substructures in A3128.  On the other hand, only the F1 and G1 groups
appear to be in a region coincident with the main X-ray activity, and
it is not even clear whether F1 and G1 are two independent systems.

Earlier we interpeted the displacement of the core emission in the NE X-ray
component from the lower surface brightness emission in terms of a ram
pressure that delineates the direction of motion of the NE peak.  In
principle, the same argument can be applied to the SW core, to infer that
its direction of motion, relative to the main cluster gas, is to the SW.
Such an inference would present serious problems for the general scenario
presented above, since we have argued that the {\it entire} F1/G1 is moving to 
the NE relative to A3128, with the SW part of it at lower speed.  However,
we propose instead that the displacement of the SW X-ray core is due to the
effect of the local gravitational potential well of a compact group centered
on the bright galaxy (ENACS\#75) that is coincident with the SW X-ray core and
the weak MOST radio source.
A CTIO 0.9-m R band image of the region immediately
surrounding ENACS\#75 is shown in Fig. \ref{cgsw}.  There it can be seen that
ENACS\#75 itself, which appears as a single bright object (at the contrast
shown) on the SuperCOSMOS grayscale
image in Fig. \ref{xrscos}, coincident with the SW X-ray core, is resolved into
three objects, with several more galaxies to the East.  The $Chandra$ SW X-ray
core component coincides with the main ENACS\#75 object.
While we have velocities only for ENACS\#75,
and for one other galaxy in the putative compact group, both of these 
(V=19252 for ENACS\#75 and V=18380 for ENACS\#78) are consistent with the
low velocity end of F1 and the mean velocity of G1.  The main point is that ENACS\#75
is displaced to the West of the compact group, thus forming a local distortion
to the potential well, and perhaps explaining the displacement of the
X-ray core.  Similarly asymmetric X-ray contours, which appear to follow the
galaxy mass distribution, have been found in small groups of galaxies (Mulchaey
\etal 1996; Davis \etal 1996; Mulchaey 2000).  The compact group also provides 
a plausible explanation for the
existence of the SW X-ray component itself.  In this scenario the SW
component reflects the nature of the ICM of an infalling group
that may be associated with the F1/G1 system, but has not yet interacted
with the A3128 ICM.  Since it has not yet passed though A3128,
its ICM is likely still bound in the group, and reflects the local structure in that potential 
well.  By way of contrast, the lack of association between the NE component and
any prominent enhancement in the galaxy distribution, as well as its more
disturbed morphology, is consistent with its identification as stripped gas from
the leading group of galaxies in the F1 filament.

\subsubsection{Comparison with Other Cluster Merging Events}

Further insight regarding the merging event(s) in A3128/A3125 can be gained by
comparing the X-ray image of A3128 to high-resolution {\it Chandra} 
imaging of other clusters with merging events.  In several clusters,  observed 
with {\it Chandra}, signatures of ongoing merging events have been established.
In particular, merging events have been studied in the clusters A2142
(Markevitch \etal 2000), A3667 (Vikhlinin, Markevitch, \& Murray 2000a), and
RXJ1720.1+2638 (Mazzotta \etal 2001).  In all three cases, the merging event
appears to consist of a subcluster ``cold'' front moving into the ICM of a
main cluster.  Due to the lack of a pressure and temperature inversion across
the front, the inference is that the infalling subcluster is moving 
subsonically.  This represents a marked contrast to the situation in A3128,
where we see no evidence that the temperature drop sin the NE and SW components,
and where we have inferred that the gravitational influence of the H-R
Supercluster has produced hypersonic infall velocities.  Thus the A3128/A3125
system potentially represents a different regime of merger activity.  A more
extensive temperature map of the cluster would be very useful in comparing with
the ``cold'' front clusters.

\subsubsection{Alternate Scenario}

Thus far we have assumed that all of the diffuse X-ray emission seen in the
$Chandra$ ACIS-I image is due to gas associated with the merging events in
A3128/A3125.  The spatial coincidence between the gas in the SW component
and the compact galaxy group in that area, coupled with the coincidence between
the peak in the SW emission and the luminous galaxy ENACS\#75 provides 
circumstantial evidence that the gas in the SW component is, indeed, affiliated
with A3128.  The redshift of 0.065 found for the SW component gas (see \S 4)
provides conclusive proof of its association with the cluster complex.  As well,
the redshift of 0.059 found for the diffuse component gas confirms its 
association with A3128.
In the case of the NE component, no such coincidence occurs with
a particular galaxy group.  The strongest circumstantial evidence for
a connection between the gas in the NE component with A3128 is the elongated
morphology of the emission, that coincides well with the position angle
of the A3128/A3125/F1 merger axis, as well as the asymmetric morphology of the
outer isophotes, which taken together suggest a direction of motion of the NE
component along the main cluster merger axis.  However, barring conclusive redshift
information on the X-ray gas, the possibility that the gas could instead be a
projection of a background cluster should be considered.  In that regard, we
again consider the nature of the radio source(s) in the NE component, and of the
background galaxy that is coincident with the double-lobed source resolved
in the ATCA 13 cm data.  As seen in the CTIO 0.9-m optical R band image, in
Fig.\ref{negal}, the background galaxy is the brightest of a number of faint
red objects in the field.  If we assume that the faint fuzz surrounding that
galaxy represents the extended envelope of a cD galaxy, i.e., that it is the
dominant galaxy of a background cluster, then by also assuming a typical
absolute magnitude for a cD of -22,
we infer a redshift of z$\sim$0.3.  The curved blue feature
to the southwest of this putative cD galaxy might then be seen as a 
gravitationally lensed arc.  Then, if the X-ray emission
in the NE is assumed to be associated with this background cluster, its
inferred luminosity is $\sim$3 x 10$^{45}$ erg/sec in the 0.5 -- 10 keV
passband.  While this is a high luminosity, it is not inconsistent with that
of the most X-ray luminous clusters known (e.g., Ebeling \etal 1996;
Wu, Xue, \& Fan 1999).  In addition, while the observed global temperature of 
the NE component, which is $\sim$4 keV, is low compared to the typical $\sim$8
keV temperature expected for such an X-ray luminous cluster, there is enough
scatter in the L$_x$ -- T relation that a $\sim$4 keV temperature is not
out of the question for the $\sim$3 x 10$^{45}$ erg/sec luminosity (Wu, Xue, \& Fan 1999).  

The key to placing the X-ray gas in the NE component is to obtain a conclusive 
measure of the redshift using the X-ray emission lines present in the spectrum. 
As was summarized in \S 4, our best fit model produces
a redshift of 0.10, with a 90\% confidence of 0.037 to 0.23.  Thus the
redshift data is more consistent with A3128, at z=0.06, than with the proposed
background cluster at z$\sim$0.3, but is not conclusive.

\section{Conclusions}

We have collected optical redshift data for 532 objects in the field of the 
merging double cluster galaxies A3128/A3125.  The redshift information has been
supplemented by both X-ray and radio imaging.  The goal of this program is to
characterize and understand the variety of structures still present in the
double cluster, and its relation to the larger environment of the 
Horologium-Reticulum Supercluster.  

Our principal conclusion is that a large number of substructures are still
present in the A3128/A3125 system.  This conclusion is based both on groups
and filaments evident in the position-velocity space of the galaxy distribution
and on the multi-component nature of the X-ray emitting ICM.  The most
striking large-scale feature in the galaxy distribution is the relatively
low number density of galaxies for $\sim$4000 \kms on either side of the
mean cluster velocity at $\sim$17500 \kms.  We interpret this feature as due
to the large gravitational potential well of the H-R Supercluster.  Within
the 4000 \kms ``depleted'' zone, those galaxies that are present appear to be
members of small groups or filaments.  The latter are extended features in
position-velocity space whose statistical reality is somewhat uncertain.  We
ascribe the filaments to the tidal distension of a group after it has fallen
through the A3128/A3125 system and is emerging out the other side.  In fact,
A3125 itself shows some characteristics of a tidally disturbed system.  We
note that the filamentary structures appear to follow the main NE-SW axis
along which most merging activity appears to be taking place.  This is the
axis which connects A3125 to the larger A3128 (at a projected separation of $\sim$6 Mpc, 
and also along which the X-ray emission is split into two bright components,
at a projected separation of $\sim$1 Mpc.

We have proposed two ongoing merging events to explain the current state of
the A3128/A3125 system.  The first is the merger event between A3125 and A3128,
which is most plausibly explained if A3125 has already passed through A3128,
and is highly dispersed as a result of the passage.  The second involves the
double peaked nature of the X-ray emission.  Here we have noticed the close
spatial correspondence between one particular high velocity filament and group 
with the X-ray emitting gas.  Although the group/filament system is not
massive, in comparison with the main body of A3128, the high infall velocity
generated by the potential well of the H-R Supercluster produces a large
energy deposition in the collision between the filament and A3128, and allows
for only a modest infalling system to produce a major impact on the cluster
ICM.  Specifically, we argue that the morphology of the NE X-ray peak, along
with its coincidence with the higher velocity end of the galaxy filament,
indicates that the NE X-ray component represents the surviving ICM of the
galaxy filament that has endured a hypersonic ($\sim$Mach 6) encounter with
A3128.  The SW X-ray component appears to be the still intact ICM of a compact
group that represents the still-infalling end of the galaxy filament-group.
This gas is believed to be responding to the potential well of a compact group
of galaxies surrounding a bright galaxy.  While the details of this picture
are still quite uncertain, the key ingredient is the high encounter velocity
produced by the H-R Supercluster, which makes even the infall of a small group
an energetic and interesting event in the life of a galaxy cluster.  Further
observations which could clarify this picture include deeper X-ray observations,
to characterize the temperature structure in the ICM, deep high resolution
radio imaging to clarify the nature of the putative radio arc associated with
the NE X-ray peak, redshifts of galaxies in the apparent compact group that is
coincident with SW X-ray component, and further redshift surveys throughout 
the H-R Supercluster.

\acknowledgements
We wish to thank the staff of the AAT for acquiring the 2dF data for us under
the service observing program and for the significant help given to us in all 
phases of the observations.
The MOST is operated by the University of Sydney and supported in part by grants
from the Australian Research Council.  The Australia Telescope is funded by the
Commonwealth of Australia for operation as a national facility managed by CSIRO.
This research has been partially supported by NSF grant AST-9900720 to the
University of North Carolina.  MJH acknowledges the financial support of a
Physics \& Mathematical Physics Scholarship at the University of Adelaide.

\appendix

\section{APPENDIX}

\subsection{Infall Velocity of the NE X-ray Component}

As has been discussed in \S 6, the F1 galaxy filament (and probably associated
G1 group) represents the strongest candidate for producing the transient 
structure seen in the Chandra image of the ICM.  If we assume that F1 is indeed
responsible for the NE and SW X-ray components, then a comparison of the
projected separation between the NE end of F1 and the outer edge of the NE 
X-ray core can yield an estimate of the relative velocities of the F1 
filament and the X-ray sources.  If the X-ray structure was generated 
at the time when the F1 group first entered the A3128 cluster, then the 
angular separation is 

$ \Delta\alpha = (v_{F1} - v_{NE}) t / D_{A3128},$

\noindent where $t$ is the time elapsed since the entry into A3128,  $v_{F1}$ is the average 
velocity of F1, $v_{NE}$ is the average velocity of the leading edge of the NE X-ray core, 
and $D_{A3128}$ is the distance to the cluster.   From Figure 7, the NE ``head'' of 
F1 is separated from the leading edge of the NE X-ray source by 
$\Delta\alpha \sim$30\arcmn.

The point at which F1 first entered the A3128 cluster is uncertain because 
A3128 does not have a sharp boundary, however, it appears that the SW edge of 
the A3128 galaxy distribution (c.f. Figure 7) is about 15' beyond the SW X-ray 
lobe.  Thus, the projected angular distance traveled by the ``head'' of  F1 since
entering A3128 is 

$ \alpha_{F1} =  v_{F1} t / D_{A3128}  = 60\arcm$.

\noindent From the ratio of these two angular separations we can therefore estimate that 
the average velocity of the NE leading edge of the X-ray source is about 1/2 the
average velocity of the leading edge of the F1 group.

The radial velocity of the ``head'' of F1 is about 4000 \kms relative to the 
A3128 rest frame which implies that the radial velocity of the X-ray structure
is about 2000 \kms (again, relative to the A3128 rest frame).  Assuming a 
random inclination with respect to the plane of the sky then 
implies that the observed X-ray structure has a velocity relative to the A3128 
cluster of $\sim$3500 \kms.  This corresponds to a supersonic Mach Number, M=6, 
for a typical cluster gas temperature of 3 keV.  
This is obviously consistent with the observed axial structure in 
the X-ray source.

\subsection{Stripping of Gas from the NE X-ray Component}

We therefore hypothesize that the NE component of the A3128 X-ray source(s) 
consists of shocked gas 
which has been stripped from a group of galaxies (now seen as the F1 filament) 
that encountered and passed through A3128 at hypersonic velocity (in excess 
of 4000 \kms).   In this scenario, as the F1 group approaches the A3128 cluster
the gas held in the core of the F1 group is shocked by the encounter with the 
gas bound to A3128.  A bow shock forms and simultaneously a reverse shock 
propagates back through the F1 gas.  This shock system is responsible for
decelerating and removing the gas from the F1 group and depositing it in the 
A3128 cluster.

The velocity of the reverse shock relative to the F1 group is given by the 
usual shock jump conditions, 

$ v_{RS}^2 = ((\gamma + 1)/2\gamma)(p_{RS}/\rho_{F1}(\vec{x})) $

\noindent where $\gamma$ is the ideal gas adiabatic index, $v_{RS}$ is the 
velocity of the reverse 
shock relative to the F1 group, $p_{RS}$ is the pressure in the reverse shock and 
$\rho_{F1}(\vec{x})$ is the local density of the unshocked gas in the F1 group
at a particular point.  The region between 
the bow shock and the reverse shock, however, is nearly isobaric so the 
reverse shock pressure is

$ p_{RS} = p_{BS} = (2\gamma/(\gamma + 1)\rho_{A3128}v_{F1}^2 $

\noindent where $p_{BS}$ is the bow shock pressure, $\rho_{A3128}$ is the ambient density in Abell 
3128, and $v_{F1}$ is the incoming velocity of the F1 group (relative to the
barycenter of A3128).  As a result there is a 
simple relation between the reverse shock velocity and the incoming F1 group 
velocity, i.e.

$ v_{RS}^2 = (\rho_{A3128}/\rho_{F1}(\vec{x})) v_{F1}^2 $

\noindent If this reverse shock velocity is greater than escape velocity for 
the F1 precursor group, then the gas behind the shock front 
will be stripped from the F1 group and eventually merge with the gas in A3128.  
In that case, the initial velocity of the stripped gas relative to A3128 is 
simply the difference between the F1 group velocity and the reverse shock 
velocity, i.e.

$ v_{STRIP} = v_{F1} - v_{RS} = v_{F1} (1 - (\rho_{A3128}/\rho_{F1}(\vec{x}))^{1/2}) $

\noindent Note that the velocity of the stripped gas relative to A3128 depends 
critically on the density ratio, $\rho_{A3128}/\rho_{F1}(\vec{x})$.  
If $\rho_{A3128}/\rho_{F1}(\vec{x}) \ge 1 $
it simply means that the gas in F1 will be stopped and deposited 
on the outer perimeter of A3128 as the F1 group begins to penetrate 
($v_{STRIP} < 0$).  A negative stripping velocity indicates that the
reverse shock gas actually has a net (small) velocity opposite to $v_{F1}$.  
This mass of rapidly stripped lower density gas may contribute to the central
X-ray component, or it may be trapped in the potential well associated with the
G1 group.
If, however, there is a high density core in the 
initial F1 gas distribution, as 
would be the case for an isothermal density distribution, then the 
$\rho_{A3128}/\rho_{F1}(\vec{x})$ ratio for this high density component
would likely be less than 1.  This high density gas would still be strongly 
decelerated (and thus displaced) from the F1 group, but 
would continue to penetrate into A3128 at a high (supersonic) velocity as 
given above for $\rho_{A3128}/\rho_{F1} < 1$. Clearly, in this case
$0 < v_{STRIP} < v_{F1}$. In other words, as F1 enters 
and passes through A3128, its gas content is essentially peeled away in layers 
corresponding to the quiescent density of the gas initially in the core of F1, 
leaving a trail of stripped gas in its wake.  The above estimate that the 
initial velocity of the shocked, dense core gas represented by the NE X-ray 
component is about 1/2 of the velocity of F1 then implies that 
$\rho_{A3128}/\rho_{F1} <  1/4$ for the central (highest) density in the core.

\subsection{Ram Pressure Confinement of the NE Core Gas}

As the reverse shock passes through the core of F1, the core gas decelerates
and is displaced from the group of galaxies.  In addition, the bow shock wraps 
around the slower moving shocked gas from the core center and establishes the 
standard condition for ram pressure confinement, i.e.

$ \rho_{A3128}v(t)^2 = \rho_{CNT}C_S^2 $

\noindent where $v(t)$ is the bulk velocity of the central core gas relative
to A3128, $\rho_{CNT}\simeq4\rho_{F1}(0)$ is the density of the shocked
central gas, and $C_S$ is the sound speed of this core gas.  Note that the
initial value for $v(t)$ is equal to $v_{STRIP}$ for $\rho_{F1}(\vec{x}) =
\rho_{F1}(0)$, i.e., the central density in the unshocked gas.

A quantitative indication of ram pressure confinement is provided by 
the observed asymmetry in the NE X-ray core, with the highest surface 
brightness contours displaced toward the outer edge of the emission.
This is a standard signature of ram pressure confinement 
and deceleration.  Furthermore, the axially elongated structure of the NE core 
can be explained in the context of a ram pressure confinement model by the 
following argument.  Ram pressure confinement controls the transverse 
expansion of the X-ray source, but it does not control the axial expansion as 
the core gas backfills its trailing wake cavity at the speed of sound.  The 
ratio of these two expansion rates is 

$ \dot{L}_{TRANSVERSE} / \dot{L}_{PARALLEL} = 2\dot{R} / C_S = \xi^{-1/2} (v_{STRIP} / v(t))^{3/4} $

\noindent where $\dot{R}$ is simply the time derivative of the transverse
radius, $v_{STRIP}$ is the initial velocity of the ram pressure confined
central core gas relative to A3128, and $\xi = \rho_{CNT}/\rho_{A3128}$ is 
the density ratio.  For the NE
component, using the above estimate that $\rho_{A3128}/\rho_{F1}(0) < 1/4$ and
$\rho_{CNT}/\rho_{F1}(0) \sim 4$ we conclude that $\xi > 16$.  
Since the X-ray emitting cloud is slowing down, $v_{STRIP} / v(t)$ is 
increasing with time, so the observed value of 
$\dot{L}_{TRANSVERSE} / \dot{L}_{PARALLEL} = 1/2$ in conjunction 
with our estimate that $\xi > 16$ implies that the present velocity of
the X-ray core has been reduced to about 40\% of its initial value.

\subsection{Dynamical versus Radiative Timescales}

Neglecting radiative cooling for the moment, one can estimate the 
dynamical timescale for the X-ray components using the standard 
ram pressure confinement arguments (Christiansen 1969),

$ t_{DYN} = (8/3) \xi(R_I/v_{STRIP}) $

\noindent $R_I$ is the initial transverse radius of the X-ray core.
For the NE component, taking $R_I\sim$26 kpc and
$v_{STRIP}\sim$3500 \kms, the dynamical timescale for the NE component is

$ t_{DYN} = 2.3 \times 10^7 \xi$  yr

\noindent For the NE component, using the above estimate that 
$\rho_{A3128}/\rho_{F1}(0) < 1/4$ and 
$\rho_{CNT}/\rho_{F1}(0) \sim 4$ we conclude that $\xi > 16$ so 

$ t_{DYN} > 3.2 \times 10^8$            yr

The cooling time for the hot cluster gas is

$ t_{COOL} = (8.5 \times 10^{10}) ( {n_{P}\over 10^{-3} cm^{-3}})^{-1} 
 { T_{g} \over 10^8 K}) ^{1/2} (yr) $ 

\noindent (Sarazin 1986), where $n_{P}$ is the proton density and $T_{g}$ is the 
temperature of the gas in Kelvins. 

From the central peaks of the two components we 
estimate that the central densities of the two X-ray components are
$n_{NE} \sim 1.1 \times 10^{-2}$ and $n_{SW} \sim 2.4 \times 10^{-2}$. Thus 
cooling times in these components are $t_{NE} = 1.4 \times 10^9$ yr and
$t_{Sw} = 6.5 \times 10^8$ yr. 
Although these cooling times are greater than the dynamical timescale, they are
at most a factor of 2-4 larger, hence radiative losses may be an important 
dynamical factor in slowing the expansion of the X-ray sources as they
decelerate (c.f., Christiansen, Rolison, \& Scott 1979).

\newpage

\begin{figure}
\caption{Comparison of the deredshifted	2dF and Argus spectra of the 
galaxies \#481 and \#477 in the vicinity of the [OII]$\lambda$3727 emission 
line.  Spectra (a) and (b) are the Argus and 2dF spectra for galaxy \#481,
while (c) and (d) are 2dF and Argus spectra for galaxy \#477.  The continua
of the 2dF and Argus spectra do not match well due to the fact that the 2dF
spectra are not flux-calibrated.}
\label{emission}
\end{figure}

\begin{figure}
\caption{Dec versus RA plot for all galaxies in our A3128/A3125 sample with
redshifts in the range 9000 $<$ cz $<$ 29000 \kms, with galaxies in the more
restricted range 16000 $<$ cz $<$ 19500 \kms marked as filled squares.  
The positions are given in arcminutes from the center of A3128 ($\alpha$ = 
03:30:43.8 $\delta$ = -52:31:30 J2000).  The two bright X-ray peaks
are denoted as six-pointed stars.}
\label{cluster}
\end{figure}

\begin{figure}
\caption{Dec (top) and RA (bottom) are plotted versus redshift for our
A3128/A3125 sample.  Positions are in arcminutes from the center of A3128.}
\label{allv}
\end{figure}

\begin{figure}
\caption{Dec (top) and RA (bottom) are plotted versus redshift for our
A3128/A3125 sample.  Positions are in arcminutes from the center of A3128.
A more restricted redshift range is plotted than in Fig. \ref{allv}.  Various
foreground and background redshift slices, discussed later in the text, are
identified with different colors and symbol types.}
\label{foba_xyv}
\end{figure}

\begin{figure}
\caption{Histogram of redshifts in the A3128/A3125 field.  The major redshift
peak for the A3128/A3125 double cluster is clearly evident at $\sim$17750 \kms,
as well as the peaks at $\pm$4000 \kms of the mean A3128/A3125 redshift, and
the peak at $\sim$33000 \kms.}
\label{velhist}
\end{figure}

\clearpage

\begin{figure}
\caption{Position-velocity plots with a number of groups and filaments 
identified.  Red and blue squares are filaments F1 and F2 respectively, while 
the green and magenta unfilled circles are groups G1 and G2.}
\label{groups2_xyv}
\end{figure}

\newpage

\begin{figure}
\caption{RA versus Dec position-position plot with the groups and filaments
identified in Fig. \ref{groups2_xyv} plotted with the same symbols
and colors as in that Figure.}
\label{groups2_xy}
\end{figure}

\begin{figure}
\caption{Position-velocity plots with a statistical ``control'' group 
identified.}
\label{control_xyv}
\end{figure}

\begin{figure}
\caption{Position-position plot with the same statistical ``control'' group 
identified as in Fig. \ref{control_xyv}.}
\label{control_xy}
\end{figure}

\begin{figure}
\caption{Position-velocity plots with a number of small groups identified.
The cyan, magenta, and blue open circles are groups G3, G6, and G7, while the
green and red squares are G4 and G5.}
\label{groups3_xyv}
\end{figure}

\newpage

\begin{figure}
\caption{RA versus Dec position-position plot with the small groups 
identified in Fig. \ref{groups3_xyv} plotted with the same symbols
and colors as in that Figure.}
\label{groups3_xy}
\end{figure}

\newpage

\begin{figure}
\caption{RA versus Dec position-position plot with the different redshift slices
identified in Fig. \ref{foba_xyv} plotted with the same symbols
and colors as in that Figure.}
\label{foba_xy}
\end{figure}

\begin{figure}
\caption{Galaxy redshifts are plotted versus their radial distance from the
center of A3128.  Again, the $\sim$4000 \kms ``depletion'' zone on either side
of the mean cluster redshift can be seen.}
\label{radius}
\end{figure}

\clearpage

\begin{figure}
\caption{Chandra ACIS-I image of A3128, in the energy range 0.5 -- 10 keV,
displayed at a contrast level to
emphasize the cores of the two high surface brightness components.  The 
horizontal axis gives the full 16\arcm field of the ACIS-I image.  The image has
been binned by 4 x 4 pixels, and then smoothed with a gaussian of $\sigma$=4.0
pixels.}
\label{ds9_3}
\end{figure}

\begin{figure}
\caption{Chandra ACIS-I image of A3128, in the energy range 0.5 -- 10 keV, 
displayed at a contrast level to
emphasize the lower surface brightness features in the image, notably the
third X-ray component that lies between the two bright emission peaks.  The
horizontal axis gives the full 16\arcm field of the ACIS-I image.  The image has
been binned by 4 x 4 pixels, and then smoothed with a gaussian of $\sigma$=4.0
pixels.  The gaps between the four individual CCD I0 -- I3 chips, can be seen
in the plot, despite the dithering of the observation.}
\label{ds9_2}
\end{figure}


\begin{figure}
\caption{Adaptively smoothed Chandra image of A3128, showing the 0.5 -- 10 keV
emission contours overlaid
on a grayscale representation of the image.  The displacement of the NE and SW
X-ray peaks from their fainter contours can be clearly seen.  The full 16\arcm
x 16\arcm of the ACIS-I image is shown.}
\label{adapt_1}
\end{figure}

\begin{figure}
\caption{Contours from the $Chandra$ ACIS-I image overlaid on a grayscale 
representation of the SuperCOSMOS digitized image of the B$_J$ bandpass UK 
Schmidt plate. The $Chandra$ contour levels are 0.3, 0.4, 0.5, 0.7, and 1.0
counts/sec in the smoothed image.}
\label{xrscos}
\end{figure}


\begin{figure}
\caption{Radial surface brightness profiles are plotted for both the NE (top)
and SW (bottom) X-ray peaks versus Radius (left panels) and log(R) (right
panels).}
\label{profs}
\end{figure}

\begin{figure}
\caption{X (left panels) and Y (right panels) centers of the ellipses fitted 
to the X-ray surface brightness in the NE (top) and SW (bottom) components.
The X-Y coordinates, which are RA and Dec respectively, are given in pixels,
where each pixel corresponds to 2''.}
\label{centers}
\end{figure}

\begin{figure}
\caption{Ellipticity (left panels) and position angle (right panels) profiles
are plotted versus semi-major axis for the NE (top) and SW (bottom) components.}
\label{ellpa}
\end{figure}

\clearpage

\begin{figure}
\caption{Contours of the MOST 843 MHz radio emission are overlaid on a 
grayscale representation of the Chandra ACIS-I image of A3128.  The MOST 
contour levels are 2, 3, 5, 7, 10, 15, 20, 30, 50, 70, and 100 mJy/beam, where
the beamsize is 54\arcs x 43\arcs at position angle 0\arcdeg .}
\label{xrmost}
\end{figure}


\begin{figure}
\caption{Contours of the MOST 843 MHz radio emission are overlaid on a
grayscale representation of the optical SuperCOSMOS digital B$_J$ image.  MOST
contour levels are the same as in Fig. \ref{xrmost}.}
\label{mostscos}
\end{figure}

\begin{figure}
\caption{CTIO 0.9-m telescope images in the B (left) and R (right) bandpasses
of the faint galaxy associated with the radio source in the NE X-ray component.
Exposure times for both images were 15 minutes.  The field of view is 2\arcmn
x 2\arcmn.  The curved feature to the southwest of the galaxy is most evident
in the B image.  North is to the top and East to the left.}
\label{negal}
\end{figure}

\begin{figure}
\caption{Radio contours from the ATCA data are overplotted on the CTIO 0.9m
optical R band image.  The 20cm (blue) and 13cm (red) contours are coincident 
with a faint galaxy.  The higher resolution 13cm data clearly show the double 
lobed structure of the radio source.  Both radio images use uniform weighting.
The 20cm beam is 10.2\arcs x 6.3\arcs (PA -11\arcdeg ) and the 13cm beam is 
5.2\arcs x 3.2\arcs (PA -11\arcdeg ).  The 20 cm contour levels are 0.5,
1, 2, 4, and 8 mJy/beam.  The 13 cm contour levels are 2, 4, 6, and 8 mJy/beam.}
\label{atca_20_13}
\end{figure}

\begin{figure}
\caption{MOST (green) and ATCA 20 cm (blue) radio contours and $Chandra$ X-ray 
(red) contours of the NE X-ray peak are overlaid on the CTIO 0.9-m 
R band image of A3128.  The beam size and orientation of the 20 cm image, which
is 15.0\arcs x 9.95\arcs (PA -20\arcdeg ),
is shown as the black ellipse in the lower right corner of the figure. The
ATCA contour levels are 1, 2, 4, 8, and 16 mJy/beam.  The MOST contours
are 2, 3, 4, 7, 10, 20, and 40 mJy/beam, while the beamsize is 54\arcs x 
43\arcs (PA 0\arcdeg ).  The $Chandra$
contours are 0.14, 0.16, 0.22, 0.32, 0.45, 0.63, 0.85, 1.10, 1.39, 1.73, and
2.10 counts/sec in the smoothed image.}
\label{overlay_ne}
\end{figure}

\begin{figure}
\caption{Histogram of all redshifts in the A3128/A3125 field from 5000 \kms out 
to 65000 \kms.  The distribution in redshift is evidently nonuniform out to the
limit of the sample.}
\label{hist200_all}
\end{figure}


\begin{figure}
\caption{CTIO 0.9-m R band image of the $\sim$2.5\arcm region surrounding the
bright galaxy ENACS\#75, showing the apparent compact group of galaxies that
lies mostly to the east of it.  ENACS\#75 resolves into the triple object in 
the center of the image.}
\label{cgsw}
\end{figure}

\clearpage

\begin{deluxetable}{cccccr}
\tablenum{1}
\tablecaption{Summary of the Radio Observations \label{radio}}
\tablewidth{0pt}
\tablehead{
\colhead{Date} & \colhead{Frequency}   & \colhead{Config.}   &
\colhead{Baselines} & \colhead{Pointing centre} & \colhead{$t_{\rm obs}$}\\
\colhead{} & \colhead{(MHz)} & \colhead{} & \colhead{(m)} & \colhead{(J2000)} & \colhead{(hr)}}
\startdata
1991 Dec 03 & 843       & MOST & 15--1600 & 03 30 12.4 $-$52 33 48 & 12.0\\
1999 Feb 28 & 1384/2496 & 6C & 153--6000 & 03 31 15.0 $-$52 41 48 & 1.67 \\
2000 Mar 23 & 1384/2496 & 6D & 77--5878 & 03 31 15.0 $-$52 41 48 & 1.15 \\
2000 Nov 04 & 1384/2496 & 6D & 77--5878 & 03 31 15.0 $-$52 41 48 & 0.65 \\
\enddata
\label{radio_obs}
\end{deluxetable}

\begin{deluxetable}{rrrrrrrrrrrrc}
\tablenum{2}
\tablecolumns{13}
\tablewidth{0pc}
\tablecaption{Velocity Data from 2dF Galaxy Spectra}
\tablehead{
\colhead{ID} & \colhead{Other ID} & \multicolumn{3}{c}{$\alpha$ (J2000)} & 
\multicolumn{3}{c}{$\delta$ (J2000)} & \colhead{B$_J$}  &
\colhead{cz} & \colhead{ref} & 
\colhead{$\sigma_{V}$} & \colhead{Emission} }
\startdata
1 & & 03 & 22 & 53.67 & -52 & 39 & 51.1 & 16.8 & 18365 & a & 103.5 & n \\
3 & & 03 & 23 & 01.99 & -52 & 58 & 42.2 & 18.3 & 26219 & e & 98.1 & y \\
5 & & 03 & 23 & 10.56 & -53 & 05 & 08.1 & 17.7 & 17682 & a & 32.4 & n \\
6 & & 03 & 23 & 15.56 & -52 & 53 & 44.1 & 17.3 & 17573 & a & 37.1 & n \\
7 & & 03 & 23 & 20.91 & -52 & 26 & 20.5 & 18.4 & 34195 & e & 68.9 & y \\
8 & & 03 & 23 & 23.45 & -53 & 08 & 09.2 & 17.8 & 9877 & e & 189.7 & y \\
9 & & 03 & 23 & 26.54 & -52 & 29 & 03.5 & 16.9 & 21977 & ae & 63.9 & y \\
10 & & 03 & 23 & 38.52 & -53 & 13 & 48.7 & 16.4 & 17951 & a & 42.4 & y \\
11 & & 03 & 23 & 38.72 & -53 & 18 & 32.7 & 18.4 & 32721 & ae & 45.9 & y \\
13 & & 03 & 23 & 41.41 & -53 & 10 & 00.5 & 18.1 & 322 & a & 48.5 & n \\
15 & & 03 & 23 & 42.75 & -52 & 42 & 52.6 & 17.0 & 23583 & a & 263.6 & n \\
16 & & 03 & 23 & 43.41 & -53 & 19 & 16.9 & 17.8 & 17859 & e & 161.1 & y \\
17 & & 03 & 23 & 44.20 & -53 & 15 & 25.1 & 17.7 & 19591 & a & 71.2 & y \\
19 & & 03 & 23 & 44.57 & -53 & 17 & 32.7 & 16.6 & 17134 & ae & 239.9 & y \\
21 & & 03 & 23 & 44.59 & -53 & 20 & 46.8 & 16.7 & 23350 & ae & 39.9 & y \\
25 & & 03 & 23 & 45.33 & -52 & 50 & 27.4 & 17.2 & 23026 & ae & 58.1 & y \\
29 & & 03 & 23 & 51.31 & -53 & 19 & 04.9 & 17.2 & 17790 & a & 132.6 & y \\
32 & & 03 & 23 & 59.84 & -52 & 43 & 26.7 & 16.4 & 21935 & ae & 63.9 & y \\
33 & & 03 & 24 & 00.13 & -52 & 34 & 23.6 & 16.9 & 119 & a & 35.6 & n \\
35 & & 03 & 24 & 01.67 & -53 & 08 & 18.5 & 16.9 & 17604 & ae & 98.5 & y \\
36 & & 03 & 24 & 04.33 & -52 & 22 & 51.4 & 17.9 & 19437 & e & 89.8 & y \\
38 & & 03 & 24 & 04.67 & -53 & 18 & 53.9 & 17.7 & 23122 & ae & 104.6 & y \\
40 & & 03 & 24 & 07.26 & -52 & 49 & 43.2 & 17.4 & 24009 & ae & 104.0 & y \\
42 & & 03 & 24 & 12.05 & -53 & 20 & 32.9 & 15.9 & 17808 & ae & 46.7 & y \\
47 & & 03 & 24 & 20.70 & -52 & 57 & 35.5 & 18.4 & 17719 & a & 95.5 & n \\
48 & & 03 & 24 & 24.35 & -52 & 42 & 48.4 & 16.3 & 18211 & ae & 82.3 & y \\
50 & & 03 & 24 & 26.47 & -52 & 38 & 04.6 & 17.4 & 21971 & ae & 97.6 & y \\
53 & & 03 & 24 & 30.52 & -52 & 49 & 35.0 & 18.2 & -130 & a & 229.8 & n \\
54 & & 03 & 24 & 33.76 & -53 & 07 & 42.5 & 18.3 & 32638 & ae & 130.0 & y \\
56 & & 03 & 24 & 35.16 & -52 & 37 & 24.5 & 17.6 & 23206 & ae & 102.2 & y \\
58 & & 03 & 24 & 38.33 & -52 & 49 & 48.9 & 18.2 & 18306 & a & 62.1 & n \\
59 & & 03 & 24 & 42.40 & -52 & 33 & 53.8 & 17.3 & 14007 & e & 84.5 & y \\
60 & & 03 & 24 & 45.31 & -53 & 25 & 45.2 & 16.2 & 17716 & ae & 44.0 & y \\
62 & & 03 & 24 & 46.45 & -53 & 02 & 09.9 & 15.6 & 18 & a & 13.0 & n \\
64 & & 03 & 24 & 47.90 & -53 & 22 & 33.0 & 17.9 & 17894 & a & 42.2 & y \\
66 & & 03 & 24 & 49.04 & -52 & 39 & 10.8 & 18.2 & 52673 & e & 109.9 & y \\
68 & & 03 & 24 & 57.48 & -53 & 23 & 10.8 & 17.9 & 16701 & ae & 68.5 & y \\
69 & & 03 & 24 & 58.47 & -53 & 24 & 58.9 & 16.3 & 17956 & ae & 44.8 & y \\
70 & & 03 & 25 & 02.21 & -52 & 08 & 40.1 & 16.8 & 21324 & a & 75.4 & n \\
71 & & 03 & 25 & 02.94 & -53 & 14 & 38.3 & 17.5 & 24084 & a & 62.6 & y \\
73 & & 03 & 25 & 03.22 & -52 & 43 & 33.2 & 18.2 & 22072 & a & 58.7 & y \\
74 & & 03 & 25 & 03.81 & -52 & 48 & 44.6 & 17.9 & 18351 & a & 126.7 & n \\
75 & & 03 & 25 & 07.82 & -53 & 19 & 10.8 & 18.3 & 17966 & ae & 122.6 & y \\
76 & & 03 & 25 & 11.28 & -52 & 33 & 12.6 & 17.7 & 18227 & ae & 48.0 & y \\
77 & & 03 & 25 & 12.75 & -52 & 17 & 11.5 & 17.3 & 23119 & ae & 57.5 & y \\
78 & & 03 & 25 & 16.36 & -53 & 35 & 38.5 & 16.9 & 17825 & ae & 69.1 & y \\
79 & & 03 & 25 & 17.04 & -52 & 51 & 44.8 & 16.3 & 13813 & ae & 68.4 & y \\
80 & & 03 & 25 & 18.49 & -53 & 21 & 28.7 & 17.0 & 17996 & ae & 182.1 & y \\
81 & & 03 & 25 & 18.99 & -52 & 43 & 42.4 & 17.5 & 21835 & a & 76.9 & y \\
82 & & 03 & 25 & 20.10 & -52 & 32 & 05.8 & 16.9 & 18181 & ae & 47.1 & y \\
83 & & 03 & 25 & 20.86 & -52 & 06 & 29.3 & 17.9 & 18175 & a & 46.7 & n \\
84 & & 03 & 25 & 22.32 & -53 & 24 & 40.9 & 17.7 & 17651 & ae & 62.8 & y \\
85 & & 03 & 25 & 22.45 & -52 & 12 & 09.2 & 17.2 & 20750 & ae & 89.7 & y \\
87 & & 03 & 25 & 27.03 & -52 & 57 & 54.4 & 17.4 & 31826 & e & 108.3 & y \\
88 & & 03 & 25 & 33.01 & -52 & 39 & 27.3 & 18.5 & 39719 & ae & 219.1 & y \\
89 & & 03 & 25 & 35.66 & -52 & 34 & 54.1 & 18.2 & 18343 & ae & 225.0 & y \\
94 & & 03 & 25 & 41.72 & -52 & 55 & 04.3 & 18.5 & 176 & a & 109.9 & n \\
95 & & 03 & 25 & 42.50 & -52 & 31 & 00.0 & 16.0 & 18901 & a & 37.7 & y \\
96 & & 03 & 25 & 44.42 & -52 & 32 & 58.2 & 16.1 & 18360 & ae & 52.1 & y \\
97 & & 03 & 25 & 45.24 & -52 & 51 & 45.4 & 18.3 & 17347 & e & 58.6 & y \\
99 & & 03 & 25 & 46.62 & -53 & 28 & 53.7 & 17.4 & 32695 & ae & 148.6 & y \\
100 & & 03 & 25 & 47.75 & -52 & 53 & 32.8 & 17.5 & 18078 & e & 76.1 & y \\
101 & & 03 & 25 & 49.68 & -53 & 23 & 03.6 & 17.0 & 19117 & ae & 111.3 & y \\
102 & & 03 & 25 & 51.06 & -52 & 16 & 22.4 & 17.5 & 21152 & a & 68.0 & n \\
104 & & 03 & 25 & 58.52 & -53 & 10 & 19.4 & 18.0 & 22853 & e & 197.2 & y \\
105 & & 03 & 26 & 00.95 & -52 & 10 & 37.7 & 17.8 & 16436 & e & 93.5 & y \\
106 & & 03 & 26 & 01.07 & -53 & 26 & 06.3 & 16.2 & 18354 & ae & 23.0 & y \\
109 & & 03 & 26 & 02.42 & -52 & 08 & 04.5 & 18.2 & 39381 & a & 177.4 & n \\
110 & & 03 & 26 & 03.59 & -52 & 35 & 55.5 & 17.7 & 18017 & e & 217.4 & y \\
113 & & 03 & 26 & 09.69 & -53 & 10 & 51.9 & 17.7 & 23206 & ae & 52.9 & y \\
114 & & 03 & 26 & 10.22 & -52 & 54 & 49.1 & 18.1 & 31781 & a & 16.7 & n \\
115 & & 03 & 26 & 10.24 & -52 & 34 & 56.9 & 16.8 & 25113 & a & 206.1 & y \\
117 & & 03 & 26 & 13.02 & -52 & 28 & 37.9 & 16.9 & 18194 & a & 43.0 & n \\
118 & & 03 & 26 & 15.69 & -53 & 20 & 45.1 & 17.0 & 19370 & a & 52.5 & n \\
119 & & 03 & 26 & 18.02 & -52 & 49 & 09.1 & 17.1 & 17390 & e & 86.4 & y \\
120 & & 03 & 26 & 18.50 & -53 & 32 & 47.3 & 15.8 & 18870 & ae & 70.3 & y \\
121 & & 03 & 26 & 21.60 & -52 & 55 & 01.3 & 18.2 & 31792 & e & 56.5 & y \\
122 & & 03 & 26 & 23.13 & -52 & 39 & 43.3 & 16.5 & 18049 & ae & 74.5 & y \\
123 & & 03 & 26 & 26.16 & -52 & 12 & 29.4 & 17.7 & 17807 & e & 164.7 & y \\
124 & & 03 & 26 & 26.41 & -52 & 30 & 24.8 & 17.7 & 17759 & a & 86.3 & n \\
126 & & 03 & 26 & 27.05 & -53 & 28 & 58.6 & 17.8 & 18330 & e & 59.2 & y \\
127 & & 03 & 26 & 30.91 & -53 & 37 & 34.9 & 17.3 & 18792 & a & 245.2 & y \\
128 & & 03 & 26 & 31.94 & -53 & 39 & 04.6 & 16.5 & 18563 & ae & 69.8 & y \\
129 & & 03 & 26 & 32.10 & -52 & 59 & 31.7 & 16.9 & 31842 & a & 76.3 & y \\
130 & & 03 & 26 & 35.35 & -53 & 20 & 07.0 & 16.9 & 19552 & ae & 111.5 & y \\
131 & & 03 & 26 & 38.53 & -52 & 52 & 46.2 & 16.8 & 18259 & ae & 119.4 & y \\
133 & & 03 & 26 & 40.18 & -52 & 39 & 35.7 & 18.5 & 19110 & e & 243.2 & y \\
134 & & 03 & 26 & 44.39 & -52 & 00 & 50.6 & 17.9 & 26506 & e & 224.9 & y \\
135 & & 03 & 26 & 45.09 & -53 & 05 & 40.8 & 18.3 & 23327 & e & 103.7 & y \\
136 & & 03 & 26 & 47.16 & -52 & 57 & 45.6 & 17.3 & -41 & a & 90.8 & n \\
137 & & 03 & 26 & 50.23 & -53 & 21 & 39.1 & 17.5 & 32354 & ae & 96.3 & y \\
138 & & 03 & 26 & 50.68 & -53 & 31 & 05.2 & 17.1 & 18729 & a & 46.6 & n \\
141 & & 03 & 26 & 56.83 & -52 & 53 & 19.4 & 17.3 & 19606 & a & 46.7 & y \\
142 & & 03 & 26 & 57.99 & -52 & 55 & 51.5 & 18.0 & 22709 & ae & 48.1 & y \\
143 & & 03 & 27 & 01.66 & -53 & 09 & 15.6 & 16.9 & 19987 & a & 58.8 & y \\
145 & & 03 & 27 & 03.80 & -53 & 28 & 23.2 & 18.0 & 18825 & e & 52.0 & y \\
146 & & 03 & 27 & 05.04 & -53 & 21 & 41.7 & 16.9 & 18693 & e & 116.2 & y \\
149 & & 03 & 27 & 08.30 & -52 & 10 & 15.3 & 17.6 & 13441 & ae & 87.2 & y \\
150 & & 03 & 27 & 08.44 & -53 & 37 & 26.3 & 17.5 & 17565 & ae & 263.8 & y \\
151 & & 03 & 27 & 10.31 & -53 & 00 & 36.4 & 15.6 & 9083 & ae & 187.9 & y \\
152 & & 03 & 27 & 10.53 & -51 & 55 & 51.5 & 18.3 & 60556 & e & 143.5 & y \\
153 & & 03 & 27 & 10.85 & -53 & 39 & 46.9 & 14.1 & 17799 & ae & 72.7 & y \\
155 & & 03 & 27 & 12.96 & -53 & 02 & 35.1 & 17.5 & 17955 & ae & 92.9 & y \\
156 & & 03 & 27 & 16.88 & -52 & 35 & 03.5 & 18.2 & 18267 & a & 52.7 & n \\
157 & & 03 & 27 & 19.06 & -52 & 47 & 35.3 & 18.1 & 24478 & a & 397.8 & n \\
158 & & 03 & 27 & 20.20 & -53 & 28 & 34.5 & 17.3 & 18476 & ae & 145.0 & y \\
159 & & 03 & 27 & 20.35 & -52 & 34 & 14.2 & 18.4 & 19512 & e & 109.8 & y \\
165 & & 03 & 27 & 27.70 & -52 & 29 & 14.9 & 17.9 & 19723 & e & 0.0 & y \\
168 & & 03 & 27 & 29.36 & -52 & 52 & 48.7 & 17.6 & 22293 & ae & 135.0 & y \\
170 & & 03 & 27 & 30.99 & -52 & 38 & 18.4 & 18.1 & 18489 & a & 162.8 & n \\
172 & & 03 & 27 & 31.09 & -53 & 23 & 40.4 & 17.6 & 19359 & a & 49.8 & n \\
174 & & 03 & 27 & 32.98 & -53 & 20 & 22.3 & 16.5 & 17866 & a & 69.8 & y \\
175 & & 03 & 27 & 33.98 & -53 & 40 & 07.3 & 17.6 & 19005 & a & 32.8 & n \\
178 & & 03 & 27 & 40.10 & -51 & 59 & 22.4 & 18.0 & 17 & a & 36.8 & n \\
180 & & 03 & 27 & 40.38 & -53 & 33 & 31.6 & 17.7 & 17834 & e & 93.7 & y \\
181 & & 03 & 27 & 41.83 & -52 & 32 & 49.5 & 17.6 & 19035 & e & 232.1 & y \\
186 & & 03 & 27 & 42.76 & -53 & 38 & 23.8 & 17.4 & 17801 & ae & 118.7 & y \\
188 & & 03 & 27 & 42.90 & -53 & 34 & 23.6 & 16.3 & 18430 & a & 63.3 & n \\
191 & & 03 & 27 & 45.63 & -52 & 10 & 41.8 & 18.0 & -20 & a & 60.5 & n \\
193 & & 03 & 27 & 47.26 & -52 & 30 & 04.1 & 17.8 & 9 & a & 13.1 & n \\
195 & & 03 & 27 & 47.91 & -52 & 31 & 21.2 & 18.0 & 18525 & a & 54.1 & n \\
197 & & 03 & 27 & 49.07 & -52 & 24 & 05.9 & 17.0 & 17924 & ae & 40.7 & y \\
203 & & 03 & 27 & 52.55 & -53 & 24 & 08.5 & 16.2 & 18405 & ae & 40.9 & y \\
204 & & 03 & 27 & 53.12 & -51 & 53 & 48.2 & 17.3 & -32 & a & 60.1 & n \\
206 & & 03 & 27 & 53.93 & -52 & 23 & 03.8 & 17.8 & 17935 & a & 34.6 & n \\
207 & & 03 & 27 & 53.98 & -53 & 35 & 10.3 & 17.8 & 17974 & ae & 559.8 & y \\
209 & & 03 & 27 & 54.18 & -53 & 46 & 57.3 & 16.7 & 16034 & a & 27.2 & n \\
217 & & 03 & 28 & 03.56 & -52 & 06 & 53.7 & 18.1 & 17264 & e & 16.9 & y \\
218 & & 03 & 28 & 03.94 & -53 & 33 & 02.2 & 18.0 & 28250 & ae & 109.9 & y \\
220 & & 03 & 28 & 04.41 & -52 & 44 & 42.2 & 18.4 & 18106 & e & 201.6 & y \\
223 & & 03 & 28 & 07.49 & -52 & 36 & 50.9 & 18.4 & 16869 & ae & 62.9 & y \\
224 & & 03 & 28 & 08.56 & -52 & 30 & 31.7 & 18.1 & 11808 & ae & 250.6 & y \\
225 & E24 & 03 & 28 & 10.28 & -52 & 48 & 59.7 & 18.3 & 17939 & e & 126.8 & y \\
226 & & 03 & 28 & 11.45 & -53 & 38 & 18.7 & 17.7 & 33310 & a & 131.1 & n \\
228 & & 03 & 28 & 12.82 & -52 & 15 & 59.3 & 18.3 & 10 & a & 137.5 & n \\
229 & & 03 & 28 & 14.03 & -53 & 38 & 34.5 & 17.4 & 33218 & ae & 201.6 & y \\
231 & & 03 & 28 & 15.31 & -52 & 42 & 01.1 & 18.3 & 18031 & ae & 53.7 & y \\
233 & & 03 & 28 & 16.51 & -52 & 13 & 57.6 & 18.4 & 47184 & a & 29.0 & n \\
235 & & 03 & 28 & 17.33 & -51 & 50 & 53.3 & 17.7 & 18572 & e & 137.2 & y \\
236 & & 03 & 28 & 17.86 & -53 & 34 & 59.3 & 18.3 & 17084 & a & 65.0 & n \\
237 & & 03 & 28 & 21.35 & -52 & 35 & 19.6 & 16.6 & -43 & a & 60.7 & n \\
240 & & 03 & 28 & 23.20 & -53 & 42 & 32.6 & 18.1 & 21534 & ae & 145.3 & y \\
241 & & 03 & 28 & 23.38 & -53 & 34 & 36.0 & 16.2 & 18880 & ae & 188.7 & y \\
242 & & 03 & 28 & 24.53 & -51 & 58 & 52.1 & 17.4 & 18834 & e & 73.1 & y \\
245 & & 03 & 28 & 27.46 & -51 & 55 & 30.8 & 17.1 & 17183 & e & 67.2 & y \\
247 & & 03 & 28 & 28.17 & -53 & 27 & 28.0 & 17.6 & 79 & a & 11.6 & n \\
248 & & 03 & 28 & 28.18 & -53 & 47 & 12.7 & 17.4 & 17896 & ae & 61.1 & y \\
251 & & 03 & 28 & 30.56 & -53 & 33 & 48.5 & 17.9 & 17730 & a & 176.6 & y \\
253 & E31,CR27a & 03 & 28 & 32.40 & -52 & 57 & 09.7 & 17.2 & 18047 & a & 174.2 & n \\
254 & E30 & 03 & 28 & 32.60 & -52 & 15 & 56.3 & 18.1 & 23130 & e & 109.1 & y \\
255 & & 03 & 28 & 33.17 & -53 & 11 & 40.5 & 17.6 & 22844 & e & 91.2 & y \\
256 & & 03 & 28 & 34.44 & -53 & 24 & 35.1 & 17.0 & 18125 & ae & 64.1 & y \\
257 & & 03 & 28 & 35.08 & -53 & 48 & 43.1 & 18.1 & 33559 & ae & 149.6 & y \\
261 & & 03 & 28 & 39.75 & -53 & 19 & 29.2 & 18.3 & 22874 & e & 241.4 & y \\
263 & & 03 & 28 & 42.33 & -53 & 29 & 13.6 & 17.7 & 18092 & a & 95.4 & n \\
264 & E33 & 03 & 28 & 44.99 & -53 & 03 & 27.4 & 17.4 & 23038 & ae & 56.6 & y \\
267 & E34,CR20a & 03 & 28 & 45.99 & -52 & 58 & 50.5 & 16.8 & 18284 & a & 265.1 & n \\
268 & & 03 & 28 & 46.69 & -52 & 38 & 23.5 & 18.1 & 17084 & a & 76.2 & n \\
270 & & 03 & 28 & 47.45 & -52 & 33 & 59.6 & 18.2 & 18002 & e & 129.5 & y \\
271 & E37 & 03 & 28 & 48.73 & -52 & 27 & 18.8 & 17.6 & 17692 & ae & 68.3 & y \\
272 & & 03 & 28 & 49.09 & -53 & 27 & 52.0 & 18.1 & 17564 & e & 206.0 & y \\
274 & E38 & 03 & 28 & 51.90 & -52 & 13 & 52.1 & 16.7 & 16664 & ae & 41.9 & y \\
276 & & 03 & 28 & 53.64 & -52 & 48 & 12.5 & 18.3 & 18235 & a & 56.9 & n \\
277 & E40 & 03 & 28 & 55.01 & -52 & 12 & 58.6 & 17.3 & 18116 & a & 145.6 & n \\
281 & & 03 & 28 & 59.76 & -53 & 22 & 20.4 & 17.8 & 23078 & e & 406.4 & y \\
285 & E46 & 03 & 29 & 05.94 & -52 & 46 & 32.1 & 17.9 & 16772 & a & 39.8 & n \\
287 & & 03 & 29 & 08.09 & -53 & 42 & 15.4 & 18.1 & 17872 & e & 108.6 & y \\
288 & & 03 & 29 & 09.45 & -52 & 48 & 52.9 & 18.3 & 17427 & e & 436.8 & y \\
290 & E51 & 03 & 29 & 16.51 & -52 & 37 & 38.1 & 15.8 & 18378 & a & 104.6 & n \\
292 & & 03 & 29 & 18.95 & -52 & 12 & 38.7 & 18.5 & 18054 & a & 68.3 & n \\
293 & & 03 & 29 & 20.75 & -52 & 47 & 58.5 & 17.8 & 24720 & e & 114.3 & y \\
294 & E54 & 03 & 29 & 22.12 & -52 & 37 & 10.3 & 15.0 & 11664 & a & 42.1 & n \\
295 & & 03 & 29 & 23.59 & -51 & 56 & 36.4 & 18.3 & 50551 & a & 228.4 & n \\
296 & & 03 & 29 & 24.32 & -53 & 16 & 08.5 & 18.5 & 12005 & ae & 160.6 & y \\
297 & & 03 & 29 & 24.73 & -52 & 33 & 02.4 & 18.1 & 89 & a & 34.2 & n \\
300 & E58,CR118a & 03 & 29 & 26.47 & -52 & 30 & 27.6 & 17.0 & 19555 & a & 39.4 & n \\
301 & & 03 & 29 & 27.11 & -52 & 40 & 54.5 & 17.3 & 31884 & ae & 49.5 & y \\
305 & & 03 & 29 & 28.56 & -52 & 42 & 36.4 & 17.8 & 16378 & ae & 92.7 & y \\
306 & E61,CR167a & 03 & 29 & 29.96 & -52 & 17 & 58.2 & 17.0 & 18076 & a & 45.1 & n \\
310 & E66 & 03 & 29 & 33.38 & -52 & 46 & 33.9 & 17.5 & 16747 & a & 62.4 & n \\
311 & & 03 & 29 & 33.82 & -53 & 45 & 50.5 & 16.8 & 24763 & a & 55.5 & y \\
312 & E68 & 03 & 29 & 34.47 & -52 & 51 & 00.2 & 17.5 & 17104 & a & 65.9 & n \\
313 & & 03 & 29 & 34.92 & -53 & 33 & 03.9 & 17.9 & 17486 & ae & 38.5 & y \\
315 & E67 & 03 & 29 & 35.39 & -52 & 12 & 52.0 & 17.9 & 17848 & a & 27.2 & y \\
316 & & 03 & 29 & 36.01 & -52 & 34 & 16.5 & 18.5 & 19163 & a & 59.6 & n \\
317 & & 03 & 29 & 36.66 & -52 & 29 & 35.0 & 18.1 & 16657 & a & 194.9 & n \\
319 & & 03 & 29 & 37.01 & -53 & 34 & 55.1 & 18.1 & 34900 & e & 108.9 & y \\
320 & E70 & 03 & 29 & 37.80 & -52 & 41 & 55.5 & 16.7 & 17152 & ae & 89.7 & y \\
321 & & 03 & 29 & 37.97 & -52 & 49 & 12.8 & 18.3 & 18472 & a & 303.2 & n \\
326 & & 03 & 29 & 41.46 & -52 & 29 & 35.4 & 17.6 & 17511 & ae & 48.3 & y \\
327 & E73 & 03 & 29 & 41.57 & -52 & 31 & 16.6 & 18.1 & 18825 & a & 72.1 & n \\
328 & & 03 & 29 & 43.49 & -53 & 15 & 34.3 & 18.5 & 22916 & e & 380.5 & y \\
330 & & 03 & 29 & 45.76 & -52 & 00 & 22.0 & 18.3 & 21270 & a & 55.5 & n \\
331 & & 03 & 29 & 46.62 & -52 & 43 & 51.9 & 18.1 & -16 & a & 60.9 & n \\
332 & & 03 & 29 & 46.70 & -52 & 37 & 05.7 & 18.4 & 18712 & a & 96.1 & n \\
335 & & 03 & 29 & 50.98 & -53 & 27 & 58.9 & 17.2 & 23004 & ae & 46.4 & y \\
337 & & 03 & 29 & 51.92 & -52 & 40 & 43.4 & 17.6 & 18383 & a & 68.7 & y \\
341 & E77 & 03 & 29 & 52.85 & -52 & 51 & 55.1 & 17.3 & 17131 & a & 45.1 & n \\
343 & & 03 & 29 & 53.54 & -53 & 35 & 33.4 & 17.7 & 27973 & a & 99.6 & n \\
348 & & 03 & 29 & 56.24 & -52 & 37 & 28.2 & 18.3 & 19454 & a & 331.5 & n \\
349 & E80 & 03 & 29 & 56.83 & -53 & 09 & 03.3 & 16.7 & 19920 & e & 201.8 & y \\
351 & & 03 & 29 & 59.78 & -53 & 26 & 47.0 & 17.5 & 17904 & ae & 62.8 & y \\
352 & & 03 & 29 & 59.88 & -52 & 38 & 13.1 & 18.3 & 19309 & a & 98.4 & n \\
353 & E81 & 03 & 30 & 00.33 & -52 & 18 & 31.8 & 18.1 & 45606 & a & 140.2 & n \\
355 & & 03 & 30 & 03.00 & -52 & 55 & 30.6 & 18.3 & 16761 & a & 98.2 & n \\
361 & E84 & 03 & 30 & 05.04 & -52 & 15 & 25.1 & 17.9 & -29 & a & 64.8 & n \\
365 & & 03 & 30 & 06.11 & -53 & 42 & 37.5 & 15.4 & 15940 & ae & 43.9 & y \\
367 & & 03 & 30 & 08.47 & -53 & 23 & 51.8 & 17.3 & 17826 & e & 122.6 & y \\
368 & & 03 & 30 & 09.11 & -53 & 18 & 47.0 & 18.2 & 22630 & e & 100.3 & y \\
370 & & 03 & 30 & 10.39 & -53 & 45 & 18.6 & 17.4 & 17684 & ae & 105.5 & y \\
371 & & 03 & 30 & 10.52 & -53 & 34 & 06.6 & 17.3 & 18419 & e & 65.3 & y \\
372 & & 03 & 30 & 10.58 & -52 & 27 & 07.6 & 17.7 & 17386 & a & 38.5 & n \\
373 & E89 & 03 & 30 & 11.13 & -52 & 14 & 51.6 & 17.7 & 44935 & e & 235.6 & y \\
374 & E90 & 03 & 30 & 11.60 & -53 & 09 & 03.5 & 18.0 & 19884 & ae & 63.4 & y \\
376 & & 03 & 30 & 12.81 & -52 & 45 & 04.9 & 18.2 & 17817 & a & 56.8 & n \\
377 & & 03 & 30 & 12.95 & -51 & 52 & 31.3 & 17.4 & 25686 & a & 236.6 & n \\
378 & & 03 & 30 & 13.18 & -53 & 47 & 03.2 & 16.5 & 15833 & e & 179.1 & y \\
380 & E94 & 03 & 30 & 13.60 & -53 & 14 & 33.1 & 17.5 & 17506 & a & 46.0 & n \\
388 & & 03 & 30 & 19.46 & -53 & 37 & 03.4 & 17.8 & 17726 & a & 159.0 & n \\
390 & E99,CR130a & 03 & 30 & 22.10 & -52 & 26 & 07.6 & 17.6 & 17739 & a & 48.5 & n \\
393 & E103 & 03 & 30 & 23.11 & -53 & 02 & 58.9 & 18.0 & 22274 & ae & 115.8 & y \\
396 & & 03 & 30 & 24.09 & -52 & 44 & 37.2 & 18.2 & 19884 & ae & 63.8 & y \\
397 & & 03 & 30 & 25.00 & -53 & 36 & 50.7 & 17.7 & 17791 & a & 111.0 & n \\
402 & & 03 & 30 & 32.24 & -52 & 38 & 49.7 & 17.6 & 19651 & a & 80.9 & n \\
403 & & 03 & 30 & 33.78 & -53 & 18 & 40.9 & 17.5 & 14208 & a & 32.4 & y \\
404 & & 03 & 30 & 33.85 & -52 & 23 & 52.6 & 18.4 & 18229 & e & 16.0 & y \\
414 & CR128a & 03 & 30 & 39.71 & -52 & 26 & 28.4 & 18.3 & 17405 & a & 127.6 & n \\
415 & & 03 & 30 & 40.99 & -53 & 15 & 34.0 & 17.7 & 17652 & ae & 189.6 & y \\
416 & & 03 & 30 & 41.11 & -52 & 24 & 46.3 & 17.9 & 18689 & a & 36.6 & n \\
422 & & 03 & 30 & 46.35 & -53 & 40 & 16.8 & 17.1 & 18152 & a & 127.2 & n \\
430 & E120 & 03 & 30 & 50.26 & -52 & 14 & 29.9 & 18.2 & 20887 & a & 48.4 & n \\
431 & & 03 & 30 & 50.47 & -52 & 18 & 56.5 & 18.4 & 18542 & ae & 72.7 & y \\
432 & E124 & 03 & 30 & 50.67 & -52 & 58 & 31.0 & 17.9 & 17968 & a & 64.4 & n \\
434 & & 03 & 30 & 51.28 & -52 & 31 & 15.1 & 17.1 & -62 & a & 83.6 & n \\
438 & & 03 & 30 & 54.01 & -53 & 40 & 36.7 & 17.8 & 15921 & a & 336.7 & n \\
440 & & 03 & 30 & 54.88 & -53 & 37 & 31.0 & 17.9 & 44269 & e & 76.0 & y \\
442 & & 03 & 30 & 55.49 & -53 & 35 & 09.7 & 18.4 & 27840 & a & 114.3 & n \\
443 & & 03 & 30 & 55.62 & -51 & 51 & 50.4 & 18.5 & 18134 & e & 265.6 & y \\
445 & E131 & 03 & 30 & 56.30 & -53 & 07 & 34.5 & 16.8 & 18538 & ae & 374.9 & y \\
447 & & 03 & 30 & 56.49 & -53 & 45 & 49.8 & 16.8 & 15704 & a & 159.0 & n \\
448 & E130 & 03 & 30 & 56.52 & -52 & 51 & 20.0 & 16.8 & 18710 & ae & 48.8 & y \\
449 & E128 & 03 & 30 & 57.02 & -52 & 24 & 32.7 & 18.4 & 18221 & a & 99.9 & n \\
451 & & 03 & 30 & 59.28 & -53 & 41 & 52.8 & 17.0 & 18147 & ae & 43.0 & y \\
456 & & 03 & 31 & 01.65 & -52 & 40 & 50.4 & 18.5 & 19036 & e & 130.2 & y \\
458 & E138 & 03 & 31 & 02.14 & -53 & 06 & 06.8 & 18.0 & 18378 & ae & 111.4 & y \\
464 & & 03 & 31 & 06.67 & -52 & 30 & 39.0 & 17.8 & 17503 & a & 43.1 & n \\
467 & E143,CR144a & 03 & 31 & 08.18 & -52 & 25 & 03.2 & 17.4 & 18502 & a & 41.7 & n \\
470 & & 03 & 31 & 08.89 & -53 & 19 & 12.3 & 18.0 & 22882 & e & 168.1 & y \\
472 & & 03 & 31 & 10.47 & -52 & 55 & 41.2 & 16.9 & 21488 & ae & 81.5 & y \\
473 & E146 & 03 & 31 & 11.06 & -52 & 10 & 33.8 & 17.8 & 18385 & a & 41.3 & n \\
475 & E147 & 03 & 31 & 11.81 & -52 & 26 & 13.8 & 17.8 & 65 & a & 41.9 & n \\
477 & E148,CR80a & 03 & 31 & 12.26 & -52 & 35 & 56.8 & 15.6 & 16946 & ae & 399.1 & y \\
478 & & 03 & 31 & 12.99 & -52 & 13 & 14.0 & 18.4 & 21332 & a & 64.9 & n \\
479 & & 03 & 31 & 13.72 & -52 & 04 & 40.2 & 16.4 & 9421 & e & 113.6 & y \\
481 & E151,CR53a & 03 & 31 & 14.96 & -52 & 41 & 48.1 & 16.0 & 19936 & ae & 63.7 & y \\
482 & E150,CR156a & 03 & 31 & 15.36 & -52 & 22 & 27.4 & 16.4 & 18070 & a & 209.0 & n \\
483 & CR105a & 03 & 31 & 15.86 & -52 & 30 & 09.9 & 18.1 & 17656 & a & 24.2 & n \\
484 & E152 & 03 & 31 & 18.27 & -52 & 38 & 17.5 & 18.0 & 24327 & e & 72.9 & y \\
485 & & 03 & 31 & 18.77 & -52 & 12 & 13.1 & 18.4 & 49771 & ae & 192.4 & y \\
486 & & 03 & 31 & 19.97 & -53 & 30 & 30.6 & 18.3 & 18038 & a & 90.9 & n \\
487 & E153 & 03 & 31 & 21.45 & -53 & 01 & 43.6 & 18.0 & 23185 & a & 68.0 & n \\
489 & & 03 & 31 & 23.49 & -53 & 27 & 50.6 & 18.5 & 23247 & a & 65.9 & n \\
490 & & 03 & 31 & 23.73 & -53 & 05 & 13.5 & 18.4 & 19651 & e & 268.5 & y \\
491 & & 03 & 31 & 23.85 & -53 & 14 & 36.4 & 17.4 & 22978 & ae & 210.6 & y \\
492 & E157 & 03 & 31 & 24.80 & -52 & 50 & 30.7 & 18.1 & 18963 & a & 122.5 & n \\
493 & E156,CR126a & 03 & 31 & 25.44 & -52 & 28 & 10.0 & 15.9 & 16755 & a & 213.0 & n \\
494 & & 03 & 31 & 25.44 & -52 & 36 & 16.1 & 17.9 & 16757 & a & 31.5 & n \\
497 & & 03 & 31 & 27.26 & -53 & 01 & 45.7 & 17.7 & 22563 & e & 85.2 & y \\
498 & & 03 & 31 & 27.33 & -52 & 30 & 05.3 & 18.0 & 17490 & a & 43.8 & n \\
501 & & 03 & 31 & 30.75 & -52 & 03 & 42.6 & 18.2 & 54493 & e & 122.9 & y \\
502 & E159 & 03 & 31 & 31.16 & -52 & 20 & 44.1 & 16.3 & 18499 & ae & 41.5 & y \\
509 & E166 & 03 & 31 & 37.77 & -53 & 00 & 17.8 & 17.0 & 23042 & ae & 74.1 & y \\
510 & & 03 & 31 & 38.99 & -52 & 06 & 47.3 & 17.6 & 24991 & e & 61.7 & y \\
512 & & 03 & 31 & 40.42 & -52 & 22 & 23.5 & 18.3 & 17718 & a & 45.9 & n \\
513 & & 03 & 31 & 41.23 & -53 & 07 & 58.8 & 18.3 & 18403 & ae & 93.2 & y \\
514 & E168 & 03 & 31 & 42.30 & -52 & 25 & 11.9 & 16.9 & 19946 & a & 37.0 & n \\
516 & & 03 & 31 & 47.75 & -52 & 09 & 25.7 & 18.4 & 57996 & a & 104.0 & n \\
517 & & 03 & 31 & 50.81 & -52 & 01 & 26.1 & 18.1 & 22997 & ae & 65.4 & y \\
519 & CR161a & 03 & 31 & 52.56 & -52 & 18 & 26.0 & 17.3 & 18140 & a & 40.4 & n \\
520 & & 03 & 31 & 52.95 & -52 & 07 & 24.1 & 16.0 & 11765 & ae & 249.8 & y \\
522 & & 03 & 31 & 53.72 & -52 & 08 & 08.9 & 18.5 & 18624 & e & 105.0 & y \\
527 & & 03 & 31 & 57.79 & -52 & 41 & 13.0 & 18.1 & 16926 & ae & 86.5 & y \\
528 & E173 & 03 & 31 & 58.02 & -52 & 58 & 16.5 & 17.9 & 50096 & a & 599.6 & n \\
530 & & 03 & 32 & 04.30 & -53 & 13 & 14.1 & 17.4 & 17908 & e & 70.2 & y \\
532 & E174 & 03 & 32 & 07.51 & -52 & 15 & 39.0 & 16.7 & 18257 & e & 76.2 & y \\
533 & E175 & 03 & 32 & 07.80 & -52 & 33 & 29.8 & 17.5 & 11849 & e & 270.8 & y \\
534 & & 03 & 32 & 08.66 & -53 & 05 & 27.9 & 18.3 & 13705 & e & 62.1 & y \\
535 & & 03 & 32 & 09.70 & -53 & 29 & 58.1 & 17.9 & 32652 & a & 111.4 & n \\
537 & E181 & 03 & 32 & 12.52 & -52 & 56 & 42.4 & 17.0 & 18126 & e & 118.2 & y \\
538 & E178 & 03 & 32 & 12.57 & -52 & 43 & 49.7 & 17.2 & 17320 & a & 62.2 & n \\
540 & E182 & 03 & 32 & 14.40 & -52 & 35 & 51.1 & 16.8 & 16926 & a & 52.9 & y \\
541 & & 03 & 32 & 15.58 & -51 & 56 & 08.2 & 18.4 & -16 & a & 47.3 & n \\
544 & & 03 & 32 & 21.03 & -52 & 19 & 23.1 & 18.1 & 17987 & ae & 70.3 & y \\
547 & E185 & 03 & 32 & 26.74 & -52 & 22 & 53.6 & 16.7 & 16760 & a & 46.3 & y \\
548 & & 03 & 32 & 28.02 & -52 & 13 & 54.9 & 17.4 & 12479 & a & 103.9 & y \\
549 & & 03 & 32 & 28.24 & -53 & 09 & 20.2 & 17.8 & 33999 & e & 198.7 & y \\
550 & & 03 & 32 & 28.70 & -52 & 44 & 30.1 & 18.2 & 18125 & a & 29.1 & n \\
551 & & 03 & 32 & 29.48 & -52 & 30 & 18.4 & 17.4 & 16744 & ae & 27.5 & y \\
552 & & 03 & 32 & 29.56 & -53 & 30 & 07.4 & 17.5 & 33018 & ae & 152.4 & y \\
553 & & 03 & 32 & 30.40 & -53 & 24 & 46.7 & 18.3 & 23307 & a & 69.3 & y \\
555 & & 03 & 32 & 30.70 & -52 & 01 & 20.0 & 17.2 & 18379 & e & 249.4 & y \\
558 & & 03 & 32 & 33.55 & -52 & 32 & 24.6 & 18.0 & 18692 & ae & 102.3 & y \\
559 & E188,CR153a & 03 & 32 & 34.66 & -52 & 21 & 16.9 & 17.2 & 18280 & a & 153.7 & n \\
560 & & 03 & 32 & 39.73 & -52 & 59 & 44.7 & 18.2 & 27763 & a & 94.4 & n \\
563 & & 03 & 32 & 42.41 & -52 & 17 & 55.4 & 18.5 & 18664 & e & 110.9 & y \\
565 & & 03 & 32 & 45.32 & -52 & 07 & 38.5 & 17.2 & 18395 & ae & 117.4 & y \\
566 & & 03 & 32 & 45.40 & -52 & 10 & 57.7 & 18.1 & 20243 & e & 117.8 & y \\
567 & & 03 & 32 & 45.92 & -52 & 33 & 31.0 & 16.7 & -9 & a & 17.9 & n \\
569 & & 03 & 32 & 47.45 & -53 & 25 & 50.7 & 16.1 & 22775 & a & 81.5 & n \\
570 & & 03 & 32 & 47.49 & -53 & 28 & 24.6 & 17.9 & 23822 & ae & 145.9 & y \\
572 & & 03 & 32 & 48.41 & -52 & 51 & 50.3 & 18.4 & 33 & a & 18.1 & n \\
574 & & 03 & 32 & 50.67 & -53 & 27 & 37.3 & 17.6 & 22548 & a & 59.5 & n \\
576 & & 03 & 32 & 55.39 & -52 & 47 & 56.8 & 17.2 & 12752 & ae & 150.8 & y \\
578 & & 03 & 32 & 57.35 & -52 & 16 & 25.2 & 18.0 & -33 & a & 278.2 & n \\
579 & & 03 & 32 & 57.65 & -52 & 54 & 05.2 & 18.2 & 27779 & ae & 152.6 & y \\
581 & & 03 & 32 & 58.46 & -52 & 15 & 16.2 & 18.5 & 38932 & ae & 110.6 & y \\
582 & & 03 & 32 & 59.93 & -53 & 03 & 31.7 & 16.1 & 11796 & ae & 131.8 & y \\
583 & & 03 & 33 & 10.22 & -53 & 29 & 27.9 & 17.7 & 22533 & a & 221.6 & n \\
587 & & 03 & 33 & 17.61 & -52 & 12 & 41.2 & 17.2 & -49 & a & 285.7 & n \\
588 & E191 & 03 & 33 & 18.61 & -52 & 27 & 59.7 & 17.2 & 17676 & a & 31.7 & n \\
589 & & 03 & 33 & 18.94 & -53 & 28 & 44.9 & 17.6 & 22761 & a & 46.6 & n \\
590 & & 03 & 33 & 19.22 & -53 & 31 & 36.4 & 17.8 & 23651 & a & 63.6 & n \\
591 & E192 & 03 & 33 & 20.12 & -52 & 28 & 43.6 & 17.5 & 17569 & ae & 38.6 & y \\
592 & & 03 & 33 & 20.40 & -52 & 01 & 59.2 & 17.5 & 25216 & ae & 60.5 & y \\
593 & & 03 & 33 & 21.83 & -52 & 33 & 09.2 & 16.2 & 13405 & ae & 412.7 & y \\
595 & E193 & 03 & 33 & 24.92 & -52 & 30 & 24.9 & 17.5 & 17572 & ae & 98.7 & y \\
596 & & 03 & 33 & 25.94 & -52 & 50 & 12.7 & 18.4 & 23014 & a & 67.5 & n \\
597 & & 03 & 33 & 28.27 & -53 & 13 & 17.5 & 18.2 & 17446 & e & 248.3 & y \\
598 & & 03 & 33 & 30.06 & -53 & 10 & 23.9 & 16.2 & 17052 & ae & 273.0 & y \\
599 & & 03 & 33 & 34.79 & -53 & 12 & 24.5 & 18.3 & 18453 & e & 102.9 & y \\
602 & & 03 & 33 & 37.63 & -52 & 16 & 25.0 & 18.5 & 53569 & e & 161.5 & y \\
603 & & 03 & 33 & 38.88 & -52 & 12 & 02.7 & 16.6 & 20306 & a & 158.7 & n \\
604 & & 03 & 33 & 45.13 & -52 & 19 & 31.1 & 18.4 & 20333 & ae & 117.7 & y \\
605 & & 03 & 33 & 46.58 & -52 & 26 & 42.6 & 17.6 & 31943 & ae & 101.4 & y \\
606 & & 03 & 33 & 49.97 & -53 & 01 & 34.4 & 18.3 & 17703 & e & 185.4 & y \\
607 & & 03 & 33 & 52.74 & -52 & 14 & 00.4 & 17.8 & 19721 & e & 354.1 & y \\
608 & & 03 & 33 & 53.05 & -52 & 11 & 07.5 & 17.2 & 19983 & ae & 70.2 & y \\
611 & & 03 & 33 & 59.66 & -52 & 10 & 10.0 & 17.4 & 44705 & a & 83.6 & n \\
612 & & 03 & 34 & 03.19 & -52 & 10 & 58.2 & 17.6 & 20734 & a & 91.6 & y \\
614 & & 03 & 34 & 11.99 & -53 & 23 & 42.1 & 17.9 & 18067 & a & 68.2 & n \\
615 & & 03 & 34 & 15.31 & -52 & 43 & 38.9 & 18.4 & 31348 & ae & 58.8 & y \\
616 & & 03 & 34 & 18.43 & -53 & 07 & 32.6 & 17.7 & 20429 & a & 57.6 & y \\
617 & & 03 & 34 & 18.65 & -53 & 11 & 44.9 & 18.3 & 16953 & e & 46.5 & y \\
618 & & 03 & 34 & 21.89 & -53 & 02 & 49.8 & 17.9 & 21710 & a & 54.0 & n \\
619 & & 03 & 34 & 24.84 & -52 & 25 & 48.3 & 18.2 & 19409 & e & 85.0 & y \\
620 & & 03 & 34 & 28.25 & -52 & 33 & 41.7 & 17.1 & 13530 & e & 223.9 & y \\
622 & & 03 & 34 & 29.79 & -53 & 00 & 16.1 & 18.4 & 44485 & e & 184.7 & y \\
623 & & 03 & 34 & 33.19 & -52 & 41 & 00.8 & 17.6 & 11553 & ae & 63.9 & y \\
625 & & 03 & 34 & 37.17 & -53 & 26 & 52.4 & 17.8 & 18479 & a & 78.6 & n \\
626 & & 03 & 34 & 39.77 & -52 & 29 & 40.9 & 17.8 & 22790 & a & 76.0 & n \\
627 & & 03 & 34 & 43.14 & -52 & 50 & 00.5 & 18.4 & 44417 & e & 241.1 & y \\
628 & & 03 & 34 & 47.64 & -52 & 58 & 30.5 & 17.5 & 20339 & a & 78.7 & y \\
629 & & 03 & 34 & 49.29 & -53 & 14 & 13.7 & 16.7 & 9245 & ae & 298.8 & y \\
635 & & 03 & 35 & 07.67 & -53 & 00 & 54.1 & 16.1 & 17371 & a & 91.0 & y \\
636 & & 03 & 35 & 11.31 & -52 & 50 & 59.5 & 18.1 & 17231 & ae & 129.6 & y \\
637 & & 03 & 35 & 17.70 & -52 & 59 & 25.2 & 17.5 & 20927 & ae & 47.6 & y \\
638 & & 03 & 35 & 26.80 & -53 & 08 & 51.0 & 16.4 & 16939 & ae & 80.5 & y \\
640 & & 03 & 35 & 34.68 & -52 & 53 & 23.1 & 16.6 & 22990 & a & 66.5 & n \\
\enddata
\label{velocities_table}
\end{deluxetable}

\begin{deluxetable}{lrrcccccrr}
\tablenum{3}
\tablecolumns{10}
\tablewidth{0pc}
\tablecaption{Groups and Filaments in A3125/A3128}
\tablehead{
\colhead{(1)} & \colhead{(2)} & \colhead{(3)} & \colhead{(4)} & \colhead{(5)} &
\colhead{(6)} & \colhead{(7)} & \colhead{(8)} & \colhead{(9)} & \colhead{(10)}\\
\colhead{Group ID} & \colhead{$n$} &
\colhead{$cz$ Limits} & \colhead{$\alpha$ Limits} &
\colhead{$\delta$ Limits} & \colhead{K-S Test} & \colhead{$\alpha-cz$ Corr.} &
\colhead{$\delta-cz$ Corr.} & \colhead{$\sigma_{\alpha}$} & 
\colhead{$\sigma_{\delta}$} 
}
\startdata

G1 & 25 & 18600 -- 19400 & -18.0 -- 10.0 & \nodata & 4.4 x 10$^{-3}$ & 0.54 & 
0.40 & 6.6 & 9.7 \\
(Control) & 24  & 18600 -- 19400 & \nodata & \nodata & \nodata & 0.91 & 0.92 & 
17.1 & 31.7 \\
G2 & 14\tablenotemark{a} & 18200 -- 19100 & \nodata & -80. -- -50. & 
7.9 x 10$^{-11}$ & 0.78 & 0.14 & 9.5 & 5.5\\
(Control) & 83 & 18200 -- 19100 & \nodata & \nodata & \nodata & 0.41 & 0.97 &
20.1 & 17.1\\
F1 & 19 & 19400 -- 20500 & \nodata & -10.0 -- 22.0 & 0.20 & 0.018 &
1.13 x 10$^{-3}$ & 26.0 & 9.5\\
(Control) & 14 & 19400 -- 20500 & \nodata & \nodata & \nodata & 
1.69 x 10$^{-3}$ & 0.59 & 27.7 & 12.9\\
F2 & 10 & 19100 -- 20000 & \nodata & -53.0 -- -22.0 & 0.097 & 0.19 & 
1.75 x 10$^{-3}$ & 20.7 & 6.9\\
(Control) & 25 & 19100 -- 20000 & \nodata & \nodata & \nodata & 0.024 & 0.13 &
22.2 & 14.4\\
\multicolumn{10}{c}{ }\\
C1 & 24 & 17500 -- 18300 & -30.0 -- -22.0 & \nodata & 0.025 & 0.94 & 0.91 &
21.5 & 1.9\\
(Control) & 118 & 17500 -- 18300 & \nodata & \nodata & \nodata & 1.00 & 1.00 &
24.2 & 28.4\\
\tablenotetext{a}{One galaxy with discrepant RA has been excluded from the
original group of 15 galaxies.}
\enddata
\label{groups_table}
\end{deluxetable}

\begin{deluxetable}{lcrccrr}
\tablenum{4}
\tablecolumns{7}
\tablewidth{0pc}
\tablecaption{Small Groups and Filaments in A3125/A3128}
\tablehead{
\colhead{Group ID} & \colhead{$n$} &
\colhead{$cz$ Limits} & \colhead{$\alpha$ Limits} & \colhead{$\delta$ Limits} & 
\colhead{$\sigma_{\alpha}$} & \colhead{$\sigma_{\delta}$} 
}
\startdata

G3 & 4 & 19700 -- 20100 & \nodata & -18.0 -- -7.0 &  4.0 & 2.2 \\
(Control) & 9  & 19700 -- 20100 & \nodata & \nodata & 24.8 & 26.5 \\
G4 & 5 & 21500 -- 22400 & \nodata & -20.0 -- 5.0 & 7.1 & 6.3 \\
(Control) & 4 & 21500 -- 22400 & \nodata & \nodata & 27.5 & 22.0 \\
G5 & 5 & 20300 -- 21800 & -2.0 -- 7.0 & \nodata & 1.5 & 18.7 \\
\nodata\tablenotemark{a} & 4 & 20300 -- 21800 & -2.0 -- 7.0 & \nodata & 1.6 & 1.9 \\
(Control) & 12 & 20300 -- 21800 & \nodata & \nodata & 27.0 & 32.5 \\
G6 & 5 & 15000 -- 16300 & \nodata & -80. -- -60. & 11.0 & 2.9 \\
(Control) & 4 & 15000 -- 16300 & \nodata & \nodata & 7.0 & 4.4 \\
G7 & 8 & 22000 -- 24100 & 12.0 -- 30.0 & \nodata & 3.0 & 13.5 \\
\nodata\tablenotemark{a} & 7 & 22000 -- 24100 & 12.0 -- 30.0 & \nodata & 2.8 & 2.3 \\
(Control) & 41 & 22000 -- 24100 & \nodata & \nodata & 27.4 & 21.0 \\
\tablenotetext{a}{One galaxy with discrepant Dec has been excluded from the
original group.}
\enddata
\label{groups3_table}
\end{deluxetable}

\begin{deluxetable}{lrccc}
\tablenum{5}
\tablecolumns{4}
\tablewidth{0pc}
\tablecaption{Temperature and Abundance Data for the ICM in A3128}
\tablehead{
\colhead{Region} & \colhead{Radius} &
\colhead{kT (keV)} & \colhead{A} & \colhead{Reduced $\chi^2$}
}
\startdata

NE (Global) & 0'' - 180'' & 3.9 ${+0.7 \atop -0.6}$ & 0.13 ${0.3 \atop 0.0}$ & 1.08 \\
NE (Small Core) & 0'' - 22.5'' & 5.1 ${+2.7 \atop -1.6}$ & 0.93 ${5.0 \atop 0.0}$ & 0.99 \\
NE (Large Core) & 0'' - 45'' & 5.2 ${+2.3 \atop -1.4}$ & 0.18 ${0.7 \atop 0.0}$ & 0.72 \\
NE (Outer Region) & 45'' - 90'' & 3.9 ${+1.3 \atop -0.9}$ & 0.17 ${0.6 \atop 0.0}$ & 1.15 \\
NE (Outer Region) & 90'' - 180'' & 3.6 ${+1.0 \atop -0.6}$ & 0.13 ${0.4 \atop 0.0}$ & 1.29 \\
\multicolumn{4}{c}{ }\\
SW (Large Core) & 0'' - 90'' & 3.3 ${+0.4 \atop -0.4}$ & 0.61 ${0.9 \atop 0.4}$ & 1.29 \\
SW (Outer Region) & 90'' - 180'' & 3.7 ${+0.7 \atop -0.5}$ & 0.34 ${0.6 \atop 0.2}$ & 0.99 \\
\enddata
\label{temps_table}
\end{deluxetable}

\end{document}